\documentclass[aps,prl,reprint,balancelastpage,nofootinbib,preprintnumbers,superscriptaddress]{revtex4-1}

\usepackage{slashed}
\usepackage{bm}
\usepackage{latexsym,amssymb,amsmath,float,url,mathrsfs}
\usepackage{latexsym}
\usepackage{graphicx}
\usepackage{epstopdf}
\usepackage{amsmath}
\usepackage{comment}
\usepackage{subfigure}
\usepackage{natbib}
\usepackage{hyperref}
\usepackage{pifont}
\usepackage{blindtext}
\usepackage{adjustbox}
\usepackage{multirow}
\usepackage{tabularx}
\usepackage[table]{xcolor}
\usepackage{orcidlink}
\usepackage{tikz}
\usetikzlibrary{tikzmark}
\usepackage{fontawesome}

\usepackage{blkarray}
\usepackage{graphicx}
\usepackage{amsfonts}
\usepackage{soul}
\usepackage{amssymb}
\usepackage{amsmath}
\usepackage{cancel}
\usepackage{tcolorbox}

\usepackage{graphics,appendix,afterpage,makecell} 

\definecolor{oucrimsonred}{rgb}{0.6, 0.0, 0.0}
\definecolor{persianblue}{rgb}{0.11, 0.22, 0.73}
\definecolor{forestgreen}{rgb}{0.13,0.35,0.13}
\definecolor{lightgray}{rgb}{0.83, 0.83, 0.83}
 \hypersetup{colorlinks, citecolor=oucrimsonred, linkcolor=black, urlcolor=oucrimsonred}
\definecolor{cornellred}{rgb}{0.7, 0.11, 0.11}
\definecolor{navyblue}{rgb}{0.0, 0.0, 0.5}
\definecolor{amethyst}{rgb}{0.6, 0.4, 0.8}
\definecolor{yellow}{rgb}{1.0, 1.0, 0.0}
\definecolor{firebrick}{rgb}{0.7, 0.13, 0.13}
\definecolor{tangerineyellow}{rgb}{1.0, 0.8, 0.0}
\definecolor{deepfuchsia}{rgb}{0.76, 0.33, 0.76}
\definecolor{amber}{rgb}{1.0, 0.75, 0.0}
\definecolor{VioletRed4}{rgb}{0.55, 0.13, .32}
\definecolor{indiagreen}{rgb}{0.07, 0.53, 0.03}
\definecolor{VioletRed4}{rgb}{0.55, 0.13, .32}
\newcommand{\be}{\begin{equation}}
\newcommand{\ee}{\end{equation}}
\newcommand{\bea}{\begin{equation} \begin{aligned}}
\newcommand{\eea}{\end{aligned} \end{equation}}

\definecolor{oucrimsonred}{rgb}{0.6, 0.0, 0.0}
\newcommand\vertarrowbox[3][6ex]{%
  \begin{array}[t]{@{}c@{}} #2 \\
  \left\uparrow\vcenter{\hrule height #1}\right.\kern-\nulldelimiterspace\\
  \makebox[0pt]{\scriptsize#3}
  \end{array}%
}

\definecolor{violachiaro}{rgb}{1,0.6,1}

\definecolor{gbcolor}{rgb}{.43,.22,.12}
 
\definecolor{gbcolor2}{rgb}{.9,.2,.6}
\definecolor{gbcolor3}{rgb}{.3,.2,.6}

\definecolor{verdechiaro}{rgb}{0.6,1,0.6}
\definecolor{giallochiaro}{rgb}{1,1,0.6}
\definecolor{bluscuro}{rgb}{0.15, 0.2, 0.9}
\definecolor{verdes}{rgb}{0.1, 0.5, 0.1}%
\definecolor{tangerineyellow}{rgb}{1.0, 0.8, 0.0}

\definecolor{americanrose}{rgb}{1.0, 0.01, 0.24}
\definecolor{cobalt}{rgb}{0.0, 0.28, 0.67}
\definecolor{brandeisblue}{rgb}{0.0, 0.44, 1.0}
\definecolor{mycolor}{rgb}{0.0, 0.0, 0.5}
\definecolor{oxfordblue}{rgb}{0.0, 0.13, 0.28}
\definecolor{azure}{rgb}{0.0, 0.5, 1.0}
\definecolor{turquoiseblue}{rgb}{0.0, 1.0, 0.94}
\newtcolorbox{mynewbox}[1]{colback=white!5!white,colframe=azure!75!black,fonttitle=\bfseries,title=#1}
\newtcolorbox{mybox}{colback=mycolor!5!white,colframe=azure!75!black}
\newtcolorbox{mynamedbox}[1]{colback=mycolor!5!white,colframe=azure!75!black,title=#1}
\definecolor{venetianred}{rgb}{0.78, 0.03, 0.08}
\newtcolorbox{mynamedbox1}[1]{colback=venetianred!5!white,colframe=venetianred!80!black,title=#1}
\newtcolorbox{mynamedbox2}[1]{colback=azure!5!white,colframe=azure!80!black,title=#1}

\newcommand{\td}{{\rm d}}
\newcommand{\Msun}{M_\odot}

\definecolor{verdes}{rgb}{0.1, 0.5, 0.1}%
\definecolor{cornellred}{rgb}{0.7, 0.11, 0.11}

\definecolor{VioletRed4}{rgb}{0.55, 0.13, .32}

\hypersetup{
     colorlinks   = true,
     citecolor    = violet,
     urlcolor     = violet,
     linkcolor    = violet}

\definecolor{ForestGreen}{rgb}{0.13, 0.55, 0.13}

\definecolor{rossocorsa}{rgb}{0.83, 0.0, 0.0}

\usepackage[normalem]{ulem}

\newcommand{\papertitle}{The recent gravitational wave observation by pulsar timing arrays \\
and primordial black holes: the importance of non-Gaussianities}

\begin{document}

\title[]{\papertitle}

\author{Gabriele Franciolini\orcidlink{0000-0002-6892-9145}}
\email{gabriele.franciolini@uniroma1.it}
\affiliation{Dipartimento di Fisica, ``Sapienza'' Universit\`a di Roma, Piazzale Aldo Moro 5, 00185, Roma, Italy}
\affiliation{INFN sezione di Roma, Piazzale Aldo Moro 5, 00185, Roma, Italy}
\author{Antonio Junior Iovino\orcidlink{0000-0002-8531-5962}}
\email{antoniojunior.iovino@uniroma1.it}
\affiliation{Dipartimento di Fisica, ``Sapienza'' Universit\`a di Roma, Piazzale Aldo Moro 5, 00185, Roma, Italy}
\affiliation{INFN sezione di Roma, Piazzale Aldo Moro 5, 00185, Roma, Italy}
\affiliation{National Institute of Chemical Physics and Biophysics, R\"avala 10, Tallinn, Estonia}
\author{Ville Vaskonen\orcidlink{0000-0003-0003-2259}}
\email{ville.vaskonen@pd.infn.it}
\affiliation{National Institute of Chemical Physics and Biophysics, R\"avala 10, Tallinn, Estonia}
\affiliation{Dipartimento di Fisica e Astronomia, Universit\`a degli Studi di Padova, Via Marzolo 8, 35131 Padova, Italy}
\affiliation{INFN sezione di Padova, Via Marzolo 8, 35131 Padova, Italy}
\author{Hardi Veerm\"ae\orcidlink{0000-0003-1845-1355}}
\email{hardi.veermae@cern.ch}
\affiliation{National Institute of Chemical Physics and Biophysics, R\"avala 10, Tallinn, Estonia}


\begin{abstract}
We study whether the signal seen by pulsar timing arrays (PTAs) may originate from gravitational waves (GWs) induced by large primordial perturbations. Such perturbations may be accompanied by a sizeable primordial black hole (PBH) abundance. We improve existing analyses and show that PBH overproduction disfavors Gaussian scenarios for scalar-induced GWs at $2\sigma$ and single-field inflationary scenarios, accounting for non-Gaussianity, at $3\sigma$ as the explanation of the most constraining NANOGrav 15-year data. This tension can be relaxed in models where non-Gaussianites suppress the PBH abundance. On the flip side, the PTA data does not constrain the abundance of PBHs.
\end{abstract}

\maketitle

\noindent\textbf{Introduction --} 

The observation of a common spectrum process in the NANOGrav 12.5-year data~\cite{NANOGrav:2020bcs} sparked significant scientific interest and led to numerous interpretations of the signal as potential a stochastic gravitational wave background (SGWB) from cosmological sources, such as first order phase transitions~\cite{NANOGrav:2021flc,Xue:2021gyq,Nakai:2020oit,DiBari:2021dri,Sakharov:2021dim,Li:2021qer,Ashoorioon:2022raz,Benetti:2021uea,Barir:2022kzo,Hindmarsh:2022awe,Gouttenoire:2023naa,Baldes:2023fsp}, cosmic strings and domain walls~\cite{Ellis:2020ena,Datta:2020bht,Samanta:2020cdk,Buchmuller:2020lbh,Blasi:2020mfx,Gorghetto:2021fsn,Buchmuller:2021mbb,Blanco-Pillado:2021ygr,Ferreira:2022zzo,An:2023idh,Qiu:2023wbs,Zeng:2023jut,King:2023cgv}, or scalar-induced gravitational waves (SIGWs) generated from primordial fluctuations~\cite{Vaskonen:2020lbd,Chen:2019xse,DeLuca:2020agl,Bhaumik:2020dor,Inomata:2020xad,Kohri:2020qqd,Domenech:2020ers,Vagnozzi:2020gtf,Namba:2020kij,Sugiyama:2020roc,Zhou:2020kkf,Lin:2021vwc,Rezazadeh:2021clf,Kawasaki:2021ycf,Ahmed:2021ucx,Yi:2022ymw,Yi:2022anu,Dandoy:2023jot,Zhao:2023xnh,Ferrante:2023bgz,Cai:2023uhc} (see also \cite{Madge:2023cak}). Consequently, observation of the common spectrum process was reported by other pulsar timing array (PTA) collaborations~\cite{Goncharov:2021oub, Chen:2021rqp, Antoniadis:2022pcn}. The recent PTA data release by the NANOGrav~\cite{NANOGrav:2023gor, NANOGrav:2023hde}, EPTA (in combination with InPTA)\,\cite{EPTA:2023fyk, EPTA:2023sfo, EPTA:2023xxk}, PPTA\,\cite{Reardon:2023gzh, Zic:2023gta, Reardon:2023zen} and CPTA\,\cite{Xu:2023wog} collaborations, shows evidence of a Hellings-Downs pattern in the angular correlations which is characteristic of gravitational waves (GW), with the most stringent constraints and largest statistical evidence arising from the NANOGrav 15-year data (NANOGrav15). The analysis of the NANOGrav 12.5 year data release suggested a nearly flat GW spectrum, $\Omega_{\rm GW}\propto f^{(-1.5, 0.5)}$ at $1\sigma$, in a narrow range of frequencies around $f=5.5$ nHz. In contrast, the recent 15-year data release finds a steeper slope, $\Omega_{\rm GW}\propto f^{(1.3, 2.4)}$ at $1\sigma$ (see Fig.~\ref{fig:fits}). Motivated by this finding, a new analysis is necessary to explore which SGWB formation mechanisms can lead to the generation of a signal consistent with these updated observations.

As reported by the NANOGrav collaboration \cite{NANOGrav:2023hfp}, an astrophysical interpretation of the signal (i.e. as SGWB emitted by SMBH mergers) requires either a large number of model parameters to be at the edges of expected values or a small number of them being notably different from standard expectations. For example, the naive $\Omega \propto f^{2/3}$ scaling predicted for GW-driven supermassive black hole (SMBH) binaries is disfavoured at $2\sigma$ by the latest NANOGrav data\,\cite{NANOGrav:2023hfp, NANOGrav:2023hvm}. However, environmental and statistical effects can lead to different predictions~\cite{Sesana:2008mz, Kocsis:2010xa, Kelley:2016gse, Perrodin:2017bxr, Ellis:2023owy, NANOGrav:2023hfp, NANOGrav:2023hvm}. Although the NANOGrav analysis indicated a preference for a cosmological explanation~\cite{NANOGrav:2023hvm}, an astrophysical origin cannot certainly be ruled out at the moment.

In this letter, we consider the possibility that the recent PTA data can be explained by the SGWB associated with large curvature fluctuations generated during inflation. The SIGWs are produced by a second-order effect resulting from scalar perturbations re-entering the horizon after the end of inflation~\cite{Tomita:1975kj, Matarrese:1993zf, Acquaviva:2002ud, Mollerach:2003nq, Ananda:2006af, Baumann:2007zm, Domenech:2021ztg}. On top of SGWBs, sufficiently large curvature perturbations can lead to the formation of primordial black holes (PBH) at horizon re-entry~\cite{Carr:1974nx, Carr:1975qj, Garcia-Bellido:1996mdl, Ivanov:1994pa, Ivanov:1997ia} (see~\cite{Carr:2020gox, Green:2020jor} for recent reviews).

In general, PTA experiments are sensitive to frequencies of the SGWB associated with the production of PBHs near the stellar mass range. The possibility of PBHs constituting all dark matter (DM) is restricted in this mass range by optical lensing~\cite{EROS-2:2006ryy, Macho:2000nvd, Zumalacarregui:2017qqd, Gorton:2022fyb, Petac:2022rio, DeLuca:2022uvz} and GW observations~\cite{Raidal:2017mfl, Ali-Haimoud:2017rtz, Raidal:2018bbj, Vaskonen:2019jpv,  Hutsi:2020sol, Franciolini:2022tfm} and accretion~\cite{Ricotti:2007au, Horowitz:2016lib, Ali-Haimoud:2016mbv, Poulin:2017bwe, Hektor:2018qqw, Hutsi:2019hlw, Serpico:2020ehh}. However, the merger events involving binary PBHs can potentially account for some of the observed black hole mergers detected by LIGO/Virgo, provided they comprise $\mathcal{O}(0.1\%)$ of DM~\cite{Vaskonen:2019jpv, DeLuca:2020jug, Raidal:2017mfl, Raidal:2018bbj, Ali-Haimoud:2017rtz, DeLuca:2020qqa, Hutsi:2020sol, Franciolini:2021tla, Clesse:2020ghq, Franciolini:2022tfm}.
Crucially, requiring no PBH overproduction strongly limits the maximum amplitude of the SIGW from this scenario, as we will see in detail.

Large primordial fluctuations are possible in a wide range of scenarios including single-field inflation with specific features in the inflaton's potential~\cite{Ballesteros:2020qam,Inomata:2016rbd,Iacconi:2021ltm,Kawai:2021edk,Bhaumik:2019tvl,Cheong:2019vzl,Inomata:2018cht,Dalianis:2018frf,Kannike:2017bxn,Motohashi:2019rhu,Hertzberg:2017dkh,Ballesteros:2017fsr,Garcia-Bellido:2017mdw,Karam:2022nym,Rasanen:2018fom,Balaji:2022rsy,Frolovsky:2023hqd,Dimopoulos:2017ged,Germani:2017bcs,Choudhury:2013woa,Ragavendra:2023ret,Cheng:2021lif,Franciolini:2023lgy,Karam:2023haj,Mishra:2023lhe,Cole:2023wyx}, the most common being a quasi-inflection-point, hybrid inflation~\cite{Garcia-Bellido:1996mdl, Bugaev:2011wy, Clesse:2015wea,Pi:2017gih,Clesse:2018ogk, Spanos:2021hpk,  Spanos:2021hpk, Tada:2023pue,Tada:2023fvd,Tada:2023fvd} and models with spectator field, i.e., the curvaton~\cite{Enqvist:2001zp,Lyth:2001nq,Sloth:2002xn,Lyth:2002my,Dimopoulos:2003ii,Kohri:2012yw,Kawasaki:2012wr,Kawasaki:2013xsa,Carr:2017edp,Ando:2017veq,Ando:2018nge,Chen:2019zza,Liu:2020zzv,Pi:2021dft,Cai:2021wzd,Liu:2021rgq,Chen:2023lou,Torrado:2017qtr,Chen:2023lou,Cable:2023lca}. Even if the models generate similar peaks in the curvature power spectrum and thus also similar SIGW spectra, they may vary in the amount of non-Gaussianity (NG) which has a notable impact on the PBH abundance. We aim to extend the analysis reported by the NANOGrav collaboration~\cite{NANOGrav:2023hvm} by performing a state-of-the-art estimate of the PBH abundance and, most importantly, by considering in detail the impact of NGs in various inflationary models predicting enhanced spectral features.

\vspace{5pt}\noindent\textbf{Scalar-induced gravitational waves --} 
Scalar perturbations capable of inducing an observable SGWB and a sizeable PBH abundance must be strongly enhanced when compared to the CMB fluctuations. In the following, we aim to be as model-independent as possible and assume ans\"atze for spectral peaks applicable for classes of models.

A typical class of spectral peaks encountered, for instance, in single-field inflation and curvaton models can be described by a \emph{broken power-law} (BPL)
\be\label{eq:PPL}
    \mathcal{P}^{\rm BPL}_{\zeta}(k)
    =A \frac{\left(\alpha+\beta\right)^{\gamma}}{\left(\beta\left(k / k_*\right)^{-\alpha/\gamma}+\alpha\left(k / k_*\right)^{\beta/\gamma}\right)^{\gamma}},
\ee
where $\alpha, \beta>0$ describe respectively the growth and decay of the spectrum around the peak. One typically has $\alpha \lesssim 4$ \cite{Byrnes:2018txb}. The parameter $\gamma$ characterizes the flatness of the peak. Additionally, in quasi-inflection-point models producing stellar-mass PBHs, we expect $\beta \gtrsim 0.5$, while for curvaton models $\beta \gtrsim 2$.
Another broad class of spectra can be characterized by a \emph{log-normal} (LN) shape
\be\label{eq:PLN}
    \mathcal{P}^{\rm LN}_{\zeta}(k)
    = \frac{A}{\sqrt{2\pi}\Delta} \, \exp\left( -\frac{1}{2\Delta^2} \ln^2(k/k_*) \right)\,
\ee
Such spectra appear, e.g., in a subset of hybrid inflation and curvaton models. We find, however, that our conclusions are only weakly dependent on the details of peak shape.

The present-day SIGW background emitted during radiation domination is gauge independent~\cite{Inomata:2019yww,DeLuca:2019ufz,Yuan:2019fwv,Domenech:2020xin} and possesses a spectrum 
\bea\label{eq:GW}
h^2 \Omega_{\rm GW} (k) 
    \!=& 
    \frac{h^2 \Omega_r }{24}\!
    \left({g_*} \over {g_*^0} \right)\!
    \left({g_{*s}} \over g_{*s}^0\right)^{-\frac{4}{3}}\!
    {\cal P}_h (k),
\eea
where $g_{*s} \equiv g_{*s} \left( T_k\right)$ and $g_* \equiv g_* \left( T_k\right)$ are the effective entropy and energy degrees of freedom (evaluated at the time of horizon crossing of mode $k$ and at present-day with the superscript $0$), 
while $h^2\Omega_r = 4.2\times  10^{-5}$ is the current radiation abundance.
Each mode $k$ crosses the horizon at the temperature $T_k$ given by the relation
\begin{equation}
    k\!=\!1.5\!\times\!  10^7 {\rm Mpc^{-1}}
    \left({g_*} \over {106.75} \right)^{1\over 2}\!\!
    \left({g_{*s}} \over 106.75\right)^{-\frac{1}{3}}\!\!
    \left (\frac{T_k}{\rm GeV}\right),
\end{equation}
while corresponding to a current GW frequency
\begin{equation}
    f = 1.6\, {\rm nHz}
    \left( \frac{k}{10^6\,{\rm Mpc}^{-1}} \right).
\end{equation}
The tensor mode power spectrum is~\cite{Kohri:2018awv,Espinosa:2018eve}
\begin{align}
 \mathcal{P}_h (k)
     = &
4 \int_1^\infty \td t \int_{0}^{1}\td s 
\left [ \frac{(t^2-1)(1-s^2)}{t^2-s^2} \right ]^2 
\nonumber \\ 
    & \qquad \quad \times 
{\cal I}_{t,s}^2 \,
 \mathcal{P}_\zeta\left(k\frac{t-s}{2}\right) 
 \mathcal{P}_\zeta\left(k\frac{t+s}{2}\right),  
 \label{eq:P_h_ts}
\end{align}
where the transfer function 
\begin{align}
{\cal I}_{t,s}^2 
    & =
\frac{288(s^2+t^2-6)^2}{(t^2-s^2)^6}
\Bigg [ \frac{\pi^2}{4} (s^2+t^2-6)^2 \Theta (t-\sqrt{3})  
\nonumber\\
    & +
\left(t^2 - s^2 - \frac{1}{2} (s^2+t^2-6) \log \left| \frac{t^2-3}{3-s^2} \right| 
\right) ^2 
\Bigg ].
\label{I_RD_osc_ave_ts}
\end{align} 
To speed up the best likelihood analysis, we assume perfect radiation domination and do not account for the variation of sound speed during the QCD era (see, for example,~\cite{Hajkarim:2019nbx, Abe:2020sqb}) which also leads specific imprints in the low-frequency tail of any cosmological SGWB \cite{Franciolini:2023wjm}. 
On top of that, cosmic expansion may additionally be affected by unknown physics in the dark sector, which can, e.g., lead to a brief period of matter domination of kination~\cite{Ferreira:1997hj,Pallis:2005bb,Redmond:2018xty,Co:2021lkc,Gouttenoire:2021jhk,Chang:2021afa}. 
Both SIGW and PBH production can be strongly affected in such non-standard cosmologies~\cite{Dalianis:2019asr,Bhattacharya:2019bvk,Bhattacharya:2020lhc,Ireland:2023avg,Bhattacharya:2023ztw,Ghoshal:2023sfa}. 

Eq.~\eqref{eq:P_h_ts} neglects possible corrections due to primordial NGs. 
This is typically justified because, contrary to the PBH abundance which is extremely sensitive to the tail of the distribution, the GW emission is mostly controlled by the characteristic amplitude of perturbations, and thus well captured by the leading order.
In general, the computation of the SGWB is dominated by Eq.~\eqref{eq:P_h_ts} and remains the in the perturbative regime if $A (3f_\text{\tiny NL}/5)^2 \ll 1$, where $f_\text{\tiny NL}$ is the coefficient in front of the quadratic piece of the expansion (see Eq.~\eqref{eq:FirstExpansion} below).
For the type of NGs considered in this work, we always remain within this limit.  
Interestingly, however, both negative and positive $f_\text{\tiny NL}$ increase the SIGW abundance, with the next to leading order correction $\Omega_\text{\tiny GW}^\text{\tiny NLO}/\Omega_\text{\tiny GW} \propto A (3f_\text{\tiny NL}/5)^2$ \cite{Cai:2018dig,Unal:2018yaa,Yuan:2020iwf,Atal:2021jyo,Adshead:2021hnm,Abe:2022xur,Chang:2022nzu,Garcia-Saenz:2022tzu,Li:2023qua} (see also \cite{Bartolo:2007vp}). We leave the inclusion of these higher-order corrections for future work.

We perform a log-likelihood analysis of the NANOGrav15 and EPTA data, fitting, respectively, the posterior distributions for $\Omega_\text{\tiny GW}$ for the 14 frequency bins reported in Ref.~\cite{NANOGrav:2023gor, NANOGrav:2023hde} and for the 9 frequency bin\,\cite{EPTA:2023fyk}, including only the last $10.3$ years of data. The results are shown in Figs.~\ref{fig:cornerBPL} and \ref{fig:cornerLN} for the BPL and LN scenarios, respectively. This analysis is simplified when compared to the one reported by PTA collaborations, which fit the PTA time delay data, modelling pulsar intrinsic noise as well as pulsar angular correlations. However, it provides fits consistent with the results of the NANOGrav~\cite{NANOGrav:2023hvm} and EPTA~\cite{EPTA:2023xxk} collaborations and thus suffices for the purposes of this letter. We neglect potential astrophysical foregrounds, by assuming that the signal arises purely from SIGWs. Around $A = \mathcal{O}(1)$ or flat low $k$ tails, the scenarios considered here are also constrained by CMB observations~\cite{Pagano:2015hma, Chluba:2012we}. However, these constraints tend to be less strict than PBH overproduction and we will neglect them here.

It is striking to see that the posterior distributions shown in Figs.~\ref{fig:cornerBPL} and \ref{fig:cornerLN} for both BPL and LN analyses indicate a rather weak dependence on the shape parameters, which are ($\alpha$,$\beta$,$\gamma$) and $\Delta$, respectively, as long as the spectra are sufficiently narrow in the IR, i.e. $\alpha \gtrsim 1.1$ and $\Delta \lesssim 2.1$ at $2\sigma$.  This is because the recent PTA data prefers blue-tilted spectra generated below frequencies of SIGW peak around $k_*$.

At small scales ($k \ll k_*$), the SIGW asymptotes to (for details, see the SM)
\be
    \Omega_{\rm GW} (k \ll k_*) \, {\propto}\,  k^3(1 + \tilde A \ln^2(k/\tilde k))\, ,
\ee
where $\tilde  A$ and $\tilde k = \mathcal{O}(k_{*})$ are parameters that depend mildly on the shape of the curvature power spectrum, see more details in the Supplementary material (SM).
The asymptotic ``causality'' tail $\Omega_{\rm GW} \propto k^3$ is too steep to fit the NANOGrav15 well, being disfavoured by over $3\sigma$. However, this tension may be relieved by QCD effects~\cite{Franciolini:2023wjm}. As a result, the region providing the best fit typically lies between the peak and the causality tail, at scales slightly lower than $k_*$ at which the spectral slope is milder. Such a milder dependence can be observed in the $k_*-A$ panel of Figs.~\ref{fig:cornerBPL} and~\ref{fig:cornerLN}, where $A$ in the $1\sigma$ region scales roughly linearly with $k_*$ indicating that $\Omega_{\rm GW}$  has an approximately quadratic dependence on $k$ in the frequency range relevant PTA experiments. Additionally, since $k_* \geq 2 \times 10^7$ at $2\sigma$, the peaks in the SIGW spectrum lie outside of the PTA frequency range. This can also be observed from Fig.~\ref{fig:fits}.

\vspace{5pt}\noindent\textbf{PBH abundance --} 
To properly compute the abundance of PBHs, two kinds of NGs need to be taken into account. Firstly, the relation between curvature and density perturbations in the long-wavelength approximation is intrinsically nonlinear~\cite{Harada:2015yda,Musco:2018rwt}
\begin{equation}
    \delta(\vec{x}, t) = -\frac{2}{3} \Phi\left(\frac{1}{a H}\right)^2 e^{-2 \zeta} \left[\nabla^2 \zeta + \frac{1}{2} \partial_i \zeta \partial_i \zeta\right] \,,
\end{equation}
where $a$ denotes the scale factor, $H$ the Hubble rate, $\Phi$ is related to the equation of state parameter $w$ of the universe. For $w$ constant, $\Phi= 3(1+w)/(5+3w)$~\cite{Polnarev:2006aa}. We have dropped the explicit $\vec{x}$ and $t$ dependence or the sake of brevity.
\begin{figure}[t]
  \centering
  \includegraphics[width=0.49\textwidth]{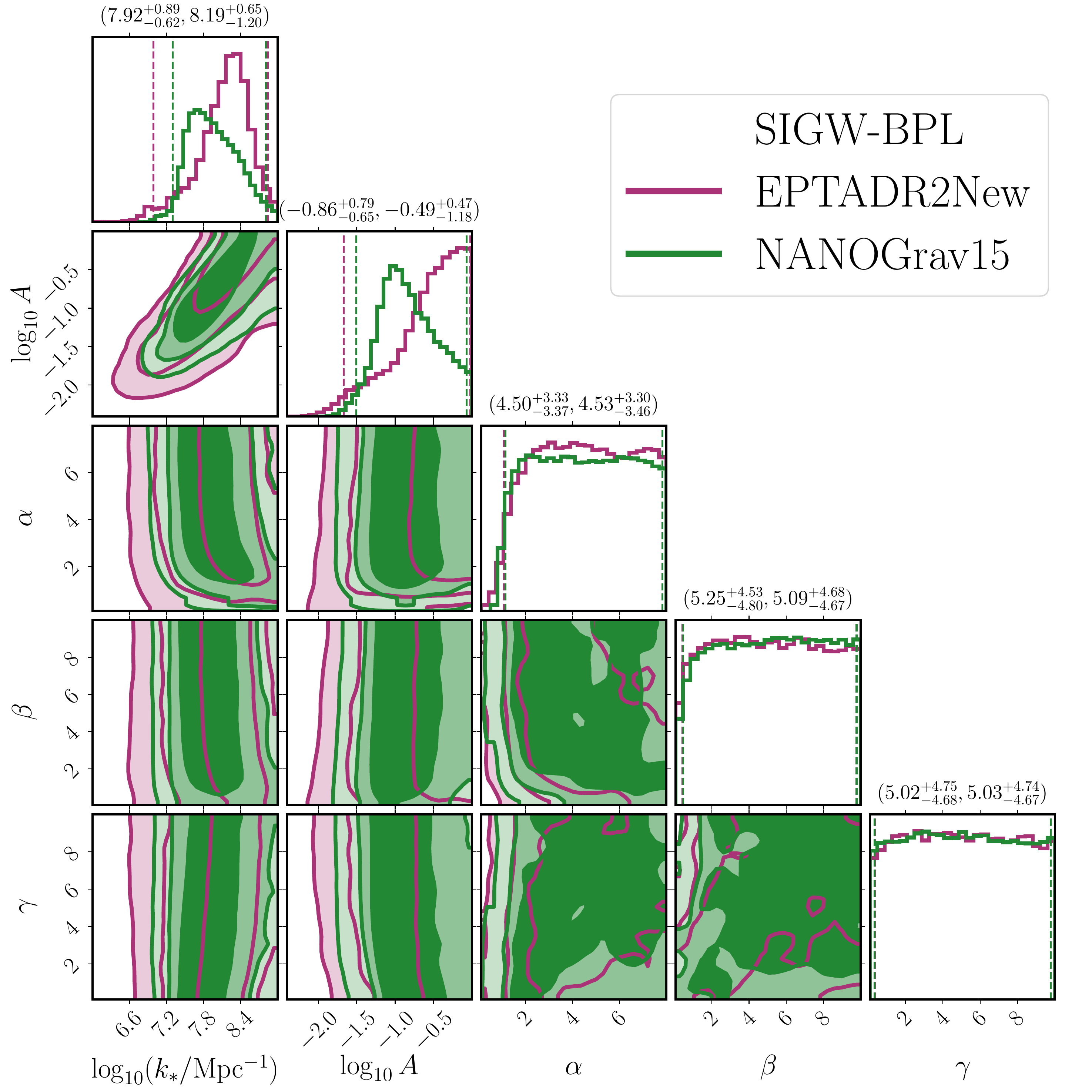}
  \caption{
  Posterior for the parameters of a BPL model \eqref{eq:PPL} for SIGWs, assuming no other source of GWs is present in both EPTA and NANOGrav15 data. 
  The shaded regions in the off-diagonal panels show 2-D posteriors at the $1\sigma$, $2\sigma$, and $3\sigma$ confidence levels and the dashed lines in the 1-D posteriors indicate the $2\sigma$ confidence level.}
  \label{fig:cornerBPL}
\end{figure}
\begin{figure}[t]
  \centering
  \includegraphics[width=0.4\textwidth]{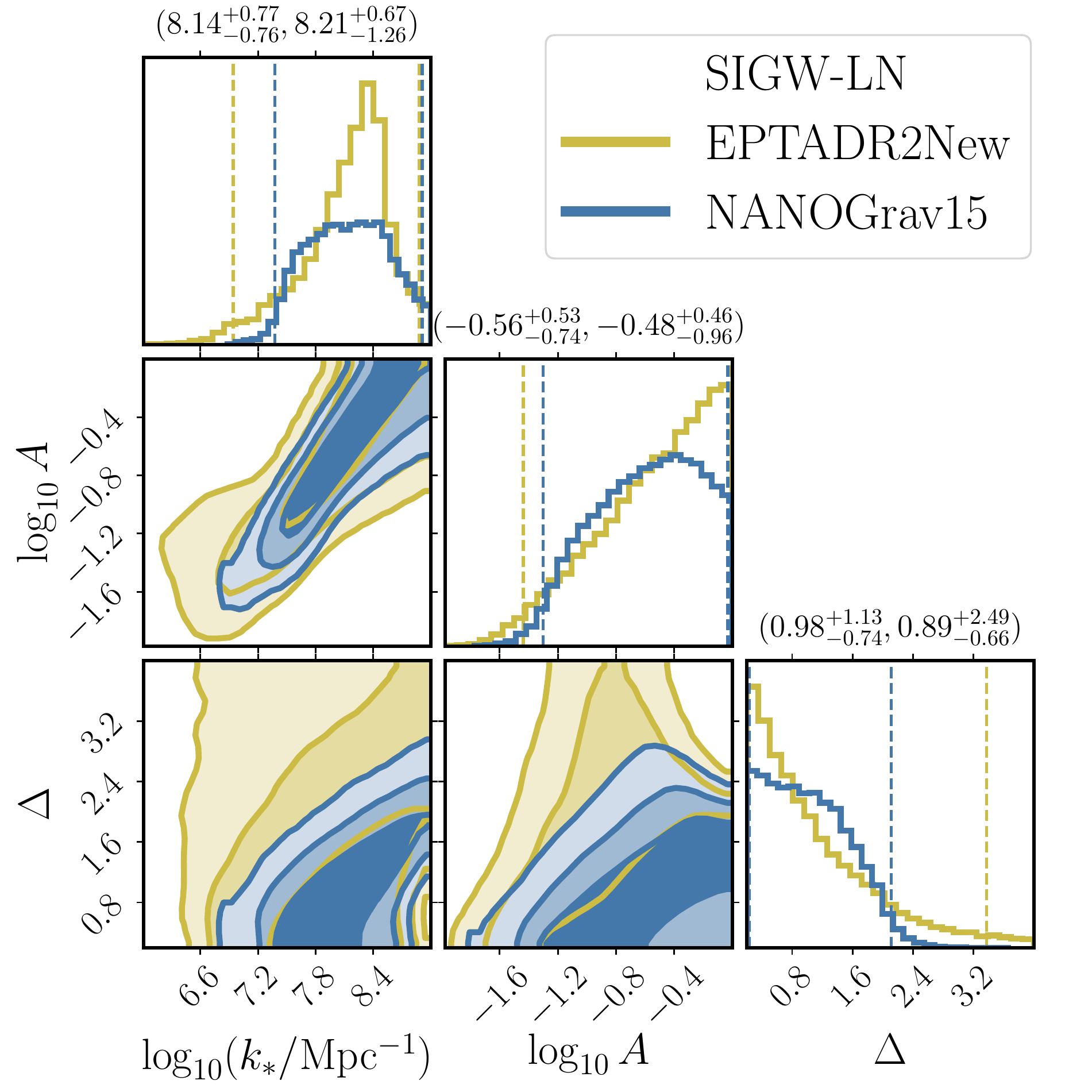}
  \caption{Same as Fig.~\ref{fig:cornerBPL}, but for the LN model \eqref{eq:PLN}.}
  \label{fig:cornerLN}
\end{figure}
Therefore, even for Gaussian curvature perturbations, the density fluctuations will inevitably inherit NGs from nonlinear corrections~\cite{DeLuca:2019qsy, Young:2019yug, Germani:2019zez}. Second, there is no guarantee that $\zeta$ is a Gaussian field -- we refer to such cases as primordial NGs. The relation  
\be\label{eq:zeta}
    \zeta(\vec{x}) = F(\zeta_{\rm G}(\vec{x}))
\ee
between $\zeta$ and its Gaussian counterpart $\zeta_G$ depends on the physical mechanism that generates the enhancement of the power spectrum at small scales. These NGs are generically independent of large-scale NGs constrained by CMB data (e.g. \cite{Planck:2019kim}).

Often, a generic model-independent approach is to consider the quadratic template
\be\label{eq:FirstExpansion}
    \zeta = \zeta_{\rm G} + \frac{3}{5}f_{\rm NL}\zeta_{\rm G}^2\,,
\ee
with $f_{\rm NL}$ as a free parameter. 
However, in explicit PBH formation models, the quadratic expansion may not be sufficient. 
Therefore, we will also consider two specific cases of $F(\zeta_{\rm G}(\vec{x}))$ in which the primordial NG can be worked out explicitly.
First, in \emph{quasi-inflection-point models} of single-field inflation, the peak in $\mathcal{P}_{\zeta}$ arises from a brief phase of ultra-slow-roll followed by constant-roll inflation dual to it~\cite{Atal:2018neu, Biagetti:2018pjj, Karam:2022nym}. In this case, the NGs can be related to the large $k$ spectral slope \cite{Atal:2019cdz,Tomberg:2023kli},
\be\label{eq:zeta_IP}
    \zeta = -\frac{2}{\beta}\log\left(1-\frac{\beta}{2}\zeta_{\rm G}\right).
\ee
Second, in \emph{curvaton models}~\cite{Sasaki:2006kq,Pi:2022ysn}, 
\be\label{eq:zeta_cur}
    \zeta = \log\big[X(r_{\rm dec},\zeta_{\rm G})\big], 
\ee
where $X(r_{\rm dec})$ is a function of $r_{\rm dec}$ (see Eq.~\eqref{eq:MasterX} in the SM for details) which we take to be the free parameter in our analysis. Curvaton self-interactions may modify the NGs (see e.g. Refs.~\cite{ Enqvist:2009ww, Fonseca:2011aa}). We omit their contribution here and leave such investigation for future work.

We follow the prescription presented in Ref.~\cite{Ferrante:2022mui} (see also \cite{Gow:2022jfb}) based on threshold statistics on the compaction function $\mathcal{C}$. The prescription improves upon the recent literature~\cite{Young:2013oia,Bugaev:2013vba,Young:2014ana,Nakama:2016gzw,Byrnes:2012yx,Franciolini:2018vbk,Yoo:2018kvb,Kawasaki:2019mbl, Riccardi:2021rlf,Taoso:2021uvl,Biagetti:2021eep,Kitajima:2021fpq,Hooshangi:2021ubn,Meng:2022ixx,Young:2022phe,Escriva:2022pnz,Hooshangi:2023kss}
by both including NL and the full primordial NG functional form \eqref{eq:zeta} non-perturbatively.\footnote{We mention here that slight discrepancies remain between peak theory and threshold statistics 
(see, e.g., Refs.~\cite{Green:2004wb,Young:2014ana,DeLuca:2019qsy}). 
As the former approach provides slightly smaller amplitudes, our conclusions remain conservative.}
The total abundance of PBHs is given by the integral (see e.g. \cite{Karam:2022nym})
\bea
    f_{\rm PBH}
&   \equiv 
    \frac{ \Omega_{\rm PBH}}{\Omega_{\rm  DM}}  
    =
    \frac{1}{\Omega_{\rm  DM}}
    \int \td \ln M_{H} \left (\frac{M_{H}}{M_\odot}\right )^{-1/2}
    \\
&   \times 
    \left({g_*} \over {106.75} \right)^{3\over 4}\!
    \left({g_{*s}} \over 106.75\right)^{-1}\! \,
    \left (\frac{\beta(M_{H})}{7.9\times 10^{-10}} \right)\,,
\eea
where $\Omega_{\rm  DM}  = 0.264$ is the cold dark matter density of the universe and the horizon mass corresponds to the temperature
\begin{equation}
    M_H(T_k) 
    = 4.8\times 10^{-2} M_\odot 
    \left(\frac{g_*}{106.75} \right)^{-{1 \over 2}} \left( T_k\over{\rm GeV} \right)^{-2}.
\end{equation}
We compute the mass fraction $\beta$ by integrating the joint probability distribution function $P_{\mathrm{G}}$ 
\be\label{eq:beta}
    \beta = 
    \int_{\mathcal{D}}\mathcal{K}(\mathcal{C} - \mathcal{C}_{\rm th})^{\gamma}
    \textrm{P}_{\rm G}(\mathcal{C}_{\rm G},\zeta_{\rm G})\td\mathcal{C}_{\rm G} \td\zeta_{\rm G}\,,
\ee
where the domain of integration is given by $\mathcal{D} =
\left\{
    \mathcal{C}(\mathcal{C}_{\rm G},\zeta_{\rm G}) > \mathcal{C}_{\rm th}  
    ~\land~\mathcal{C}_1(\mathcal{C}_{\rm G},\zeta_{\rm G}) < 2\Phi
\right\}
$, and the compaction function $\mathcal{C} = \mathcal{C}_1 - \mathcal{C}_1^2/(4\Phi)$ can be built from the linear
$\mathcal{C}_1 = \mathcal{C}_{\rm G} \, \td F/\td\zeta_{\rm G}$ component, that uses $\mathcal{C}_{\rm G} = -2\Phi\,r\,\zeta_{\rm G}^{\prime}$. The Gaussian components are distributed as
\be
    P_{\mathrm{G}}\left(\mathcal{C}_{\mathrm{G}}, \zeta_{\mathrm{G}}\right)
    = \frac{e^{\left[-\frac{1}{2\left(1-\gamma_{c r}^2\right)}\left(\frac{\mathcal{C}_{\mathrm{G}}}{\sigma_c}-\frac{\gamma_{c r} \zeta_{\mathrm{G}}}{\sigma_r}\right)^2 \!-\! \frac{\zeta_{\mathrm{G}}^2}{2 \sigma_r^2}\right]}}{2 \pi \sigma_c \sigma_\tau \sqrt{1-\gamma_{c r}^2}}.
\ee
The correlators are given by
\begin{subequations}
\begin{align}
    & \sigma_c^2=\frac{4 \Phi^2}{9} \int_0^{\infty} \frac{\td k}{k}\left(k r_m\right)^4 W^2\left(k, r_m\right)P^{T}_\zeta
    \,, \\
    &\sigma_{c r}^2=\frac{2 \Phi}{3} \! \int_0^{\infty} \!\! \frac{\td k}{k} \!\left(k r_m\right)^2 \!W\!\!\left(k, r_m\right) \!W_s\!\left(k, r_m\right) \!P^{T}_\zeta\! 
    \,, \\
    &\sigma_r^2=\int_0^{\infty} \frac{\td k}{k} W_s^2\left(k, r_m\right) P^{T}_\zeta \,,
\end{align}
\end{subequations}
with $P^{T}_\zeta=T^2\left(k, r_m\right) P_\zeta(k)$, and  $\gamma_{c r} \equiv \sigma_{c r}^2 / \sigma_c \sigma_\tau$.
We have defined $W\left(k, r_m\right), $ $W_s\left(k, r_m\right)$ and $T\left(k, r_m\right)$ 
as the top-hat window function, the spherical-shell window function, and the radiation transfer function, computed assuming radiation domination \cite{Young:2022phe}.
\footnote{The softening of the equation of state near the QCD transitions is expected to slightly affect the evolution of sub-horizon modes. Since this is mitigated by the window function that also smooths out sub-horizon modes, we neglect this effect here.}

In this work, we have followed the prescription given in Ref.~\cite{Musco:2020jjb} to compute the values of the threshold $\mathcal{C}_{\rm th}$ and the position of the maximum of the compaction function $r_m$, which depend on the shape of the power spectrum. The presence of the QCD phase transitions is taken into account by considering that $\gamma\left(M_H\right), \mathcal{K}\left(M_H\right), \mathcal{C}_{\rm th}\left(M_H\right)$ and $\Phi\left(M_H\right)$ are functions of the horizon mass around $M_{\mathrm{PBH}}=\mathcal{O}\left(\Msun\right)$~\cite{Franciolini:2022tfm,Musco:2023dak}.
We give more details in the SM.

\begin{figure}[t]
  \centering
  \includegraphics[width=0.49\textwidth]{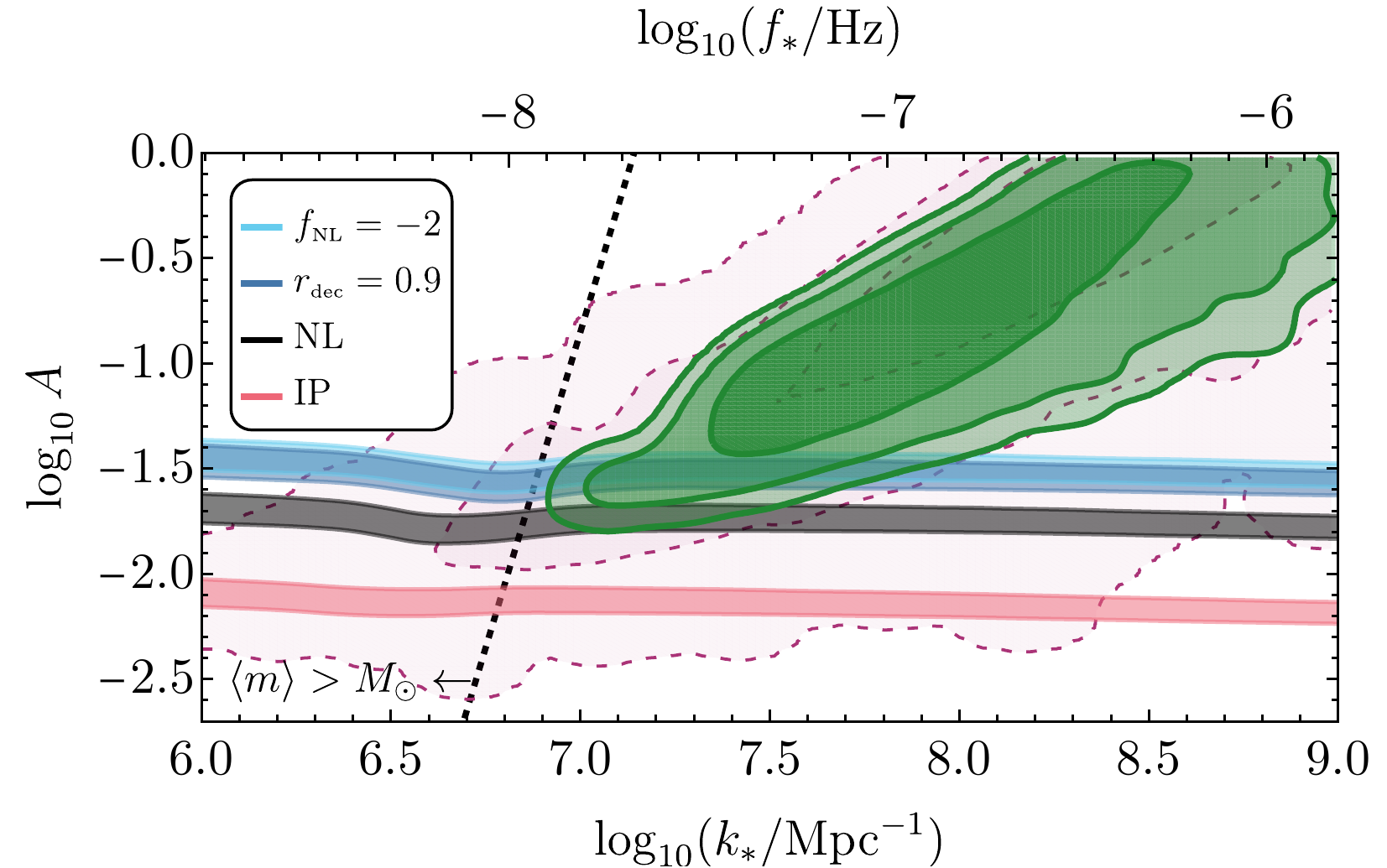}
  \caption{
   PBH abundance for different NG models: 
   non-linearities only (black), 
   quasi-inflection-point models with $\beta = 3$ (red), 
   curvaton models with $r_{\rm ec} = 0.9$ (blue) and negative $f_\text{\tiny NL}$ (cyan).
   We assume a BPL power spectrum~\eqref{eq:PPL} with $\alpha =4$, $\beta=3$ and $\gamma = 1$. 
   The colored bands cover values of PBH abundance in the range $f_{\rm PBH} \in (1,10^{-3})$ from top to bottom. 
   The green and purple posterior comes from Fig.~\ref{fig:cornerBPL}, corresponding to NANOGrav15 and EPTA, respectively. 
   The dashed line indicates an average PBH mass $\langle m \rangle = \Msun$.}
  \label{fig:abundance}
\end{figure}

The effect of NGs is illustrated in Fig.~\ref{fig:abundance} for a BPL model with $\beta=3$, $\alpha=4$, and $\gamma=1$. We find this scenario to be one of the more conservative ones, that is, changing the shape parameters or switching to an LN shape would yield similar or less optimistic conclusions for SIGW explanations of the recent PTA data.

Fig.~\ref{fig:abundance} shows that even in the absence of primordial NGs, the region avoiding overproduction of PBHs (black band and below) is excluded at over $2\sigma$ by NANOGrav15 while EPTA is currently less constraining.
This conclusion confirms the results obtained in Ref.~\cite{Dandoy:2023jot} based on IPTA-DR2 data \cite{Antoniadis:2022pcn}.
Existing constraints on the PBH abundance force $A$ to fall at the lower edge of the colored band, and slightly strengthen this conclusion. For quasi-inflection-point models, the situation is more dire as NGs tend to assist PBH production which pushes the overproduction limit below the $3\sigma$ region for NANOGrav15. Although both the slope and the NGs in the $\beta=3$ case, shown in red, are quite large, reducing the $\beta$ cannot bring these models above the black band. All in all, we can conclude that constraints on the PBH abundance disfavor quasi-inflection-point models as a potential explanation for NANOGrav15. The flip-side of this conclusion is that NANOGrav15 does not impose additional constraints on the PBH abundance. Thus, a component of the signal may be related to the formation of subsolar mass PBHs that may be independently probed by future GW experiments~\cite{DeLuca:2021hde, Pujolas:2021yaw, Miller:2020kmv, Urrutia:2023mtk, Franciolini:2023opt}.

On the other hand, the tension between SIGWs and NANOGrav15 can be alleviated in models in which NGs suppress the PBH abundance. This is demonstrated by the blue bands in Fig.~\ref{fig:abundance}, which correspond to $f_{\rm PBH} \in (10^{-3},1)$ curvaton models~\eqref{eq:zeta_cur} with a large $r_{\rm dec}$ and for a generic quadratic ansatz~\eqref{eq:FirstExpansion} with a large negative $f_{\rm NL}$. It is important to stress, that both cases displayed in Fig.~\ref{fig:abundance} represent the most optimistic scenarios: increasing $r_{\rm dec}$ above $0.9$ would have an unnoticeable effect on $f_{\rm PBH}$ and decreasing $f_{\rm NL}$ below $-2$ has a positive effect on PBH formation and would shift the lines away from the best-fit region. This is because sizeable \emph{negative} curvature fluctuations can still generate large fluctuations in the compaction and seed sizeable abundance~\eqref{eq:beta} (see the SM for further details). 

The best-fit region for NANOGrav15 lies at scales $k_*>10^7 {\rm Mpc}^{-1}$ which corresponds to the production of sub-solar mass PBHs (see Fig.~\ref{fig:abundance}). Around $k_* \approx 10^{7} {\rm Mpc}^{-1}$, small dents in the colored bands in Fig.~\ref{fig:abundance} can be observed. These arise due to the effect of the QCD phase transition which promotes PBH formation. Thus, we find that the QCD-induced enhancement of $f_{\rm PBH}$ in the parameter space relevant for NANOGrav15 tends to be negligible.

Although our $f_{\rm PBH}$ estimates assume quite narrow curvature power spectra, we checked that our conclusions about PBH overproduction in single-field inflation persist also in the case of broad spectra (e.g. see the models in Refs.~\cite{DeLuca:2020agl,Sugiyama:2020roc,Franciolini:2022pav,Ferrante:2023bgz} connecting PTA observations to asteroidal mass PBH dark matter).

As a last remark, limiting our analysis to the absence of NGs in the curvature perturbation field $\zeta$, we have found that our results differ from those published by the NANOGrav collaboration~\cite{NANOGrav:2023hvm}. These discrepancies arise because their analysis is subject to a few simplifications: the omission of critical collapse and the nonlinear relationship between curvature perturbations and density contrast, the adoption of a different value for the threshold (independently from the curvature power spectrum), and the use of a Gaussian window function (which is incompatible with their choice of threshold \cite{Young:2019osy}). Another minor limitation is that they disregard any corrections from the QCD equation of state, although we find that the result is minimally dependent on this aspect.\footnote{{\it Note added:} Similar and other simplifications were made in Ref.~\cite{Inomata:2023zup,Wang:2023ost,Liu:2023ymk} which appeared briefly after the submission of this Letter.}

\vspace{5pt}\noindent\textbf{Conclusions and outlook --} 
The evidence for the Hellings-Downs angular correlation reported by the NANOGrav, EPTA, PPTA, and CPTA collaborations sets an important milestone in gravitational-wave astronomy. One of the most pressing challenges to follow is to determine the nature signal: is it astrophysical or cosmological? 

In this letter, we have analyzed the possibility that this signal may originate from GWs induced by high-amplitude primordial curvature perturbations. This scenario is accompanied by the production of a sizeable abundance of PBHs. 
Our findings demonstrate that PBH formation models that feature Gaussian primordial perturbations, or positive NGs would overproduce PBHs unless the amplitude of the spectrum is much smaller than required to explain the GW signal.
For instance, most models relying on single-field inflation featuring an inflection point appear to be excluded at $3\sigma$ as the sole explanation of the NANOGrav 15-year data. 
However, this tension can be alleviated for models where large negative NGs suppress the PBH abundance. For instance, curvaton scenarios with a large $r_{\rm dec}$ and models exhibiting only large negative $f_{\rm NL}$. As a byproduct, however, we conclude that the PTA data does not impose constraints on the PBH abundance.

Several future steps should be taken to improve the analysis of this paper. For instance, it would be important to fully include the impact of NGs and the variation of sound speed during the QCD era when calculating the present-day SIGW background, which provides a significant computational challenge.
Beyond that, it would be important to include NGs corrections to the threshold for collapse and to reduce remaining uncertainties in the computation of the abundance. Finally, we expect that a comprehensive joint analysis involving all collaborations within the International Pulsar Timing Array (IPTA) framework will further strengthen the constraints discussed in this work.

\begin{acknowledgments}
\vspace{5pt}\noindent\emph{Acknowledgments --}
We thank V. De Luca, G. Ferrante, D. Racco, A. Riotto, F. Rompineve, A. Urbano, and J. Urrutia for useful discussions.
G.F. acknowledges the financial support provided under the European Union's H2020 ERC, Starting Grant agreement no.~DarkGRA--757480 and under the MIUR PRIN programme, and support from the Amaldi Research Center funded by the MIUR program ``Dipartimento di Eccellenza" (CUP:~B81I18001170001). This work was supported by the EU Horizon 2020 Research and Innovation Programme under the Marie Sklodowska-Curie Grant Agreement No. 101007855 and additional financial support provided by ``Progetti per Avvio alla Ricerca - Tipo 2", protocol number AR2221816C515921. 
A.J.I. acknowledges the financial support provided under the ``Progetti per Avvio alla Ricerca Tipo 1", protocol number AR12218167D66D36, and the ``Progetti di mobilità di studenti di dottorato di ricerca".
The work of V.V. and H.V. was supported by European Regional Development Fund through the CoE program grant TK133 and by the Estonian Research Council grants PRG803 and PSG869. The work of V.V. has been partially supported by the European Union's Horizon Europe research and innovation program under the Marie Sk\l{}odowska-Curie grant agreement No. 101065736.
\end{acknowledgments}

\bibliography{main}

\clearpage
\newpage
\maketitle
\onecolumngrid
\begin{center}
\textbf{\large \papertitle} 
\\ 
\vspace{0.05in}
{Gabriele Franciolini, Antonio Junior Iovino, Ville Vaskonen, and Hardi Veerm\"ae}
\\ 
\vspace{0.05in}
{ \it Supplementary Material}
\end{center}
\onecolumngrid
\setcounter{equation}{0}
\setcounter{figure}{0}
\setcounter{section}{0}
\setcounter{table}{0}
\setcounter{page}{1}
\makeatletter
\renewcommand{\theequation}{S\arabic{equation}}
\renewcommand{\thefigure}{S\arabic{figure}}
\renewcommand{\thetable}{S\arabic{table}}

\vspace{-0.5cm}
\section{Asymptotics of the GW power spectrum}\label{supp:1}

\begin{figure}[h]
  \centering
  \includegraphics[width=0.95\textwidth]{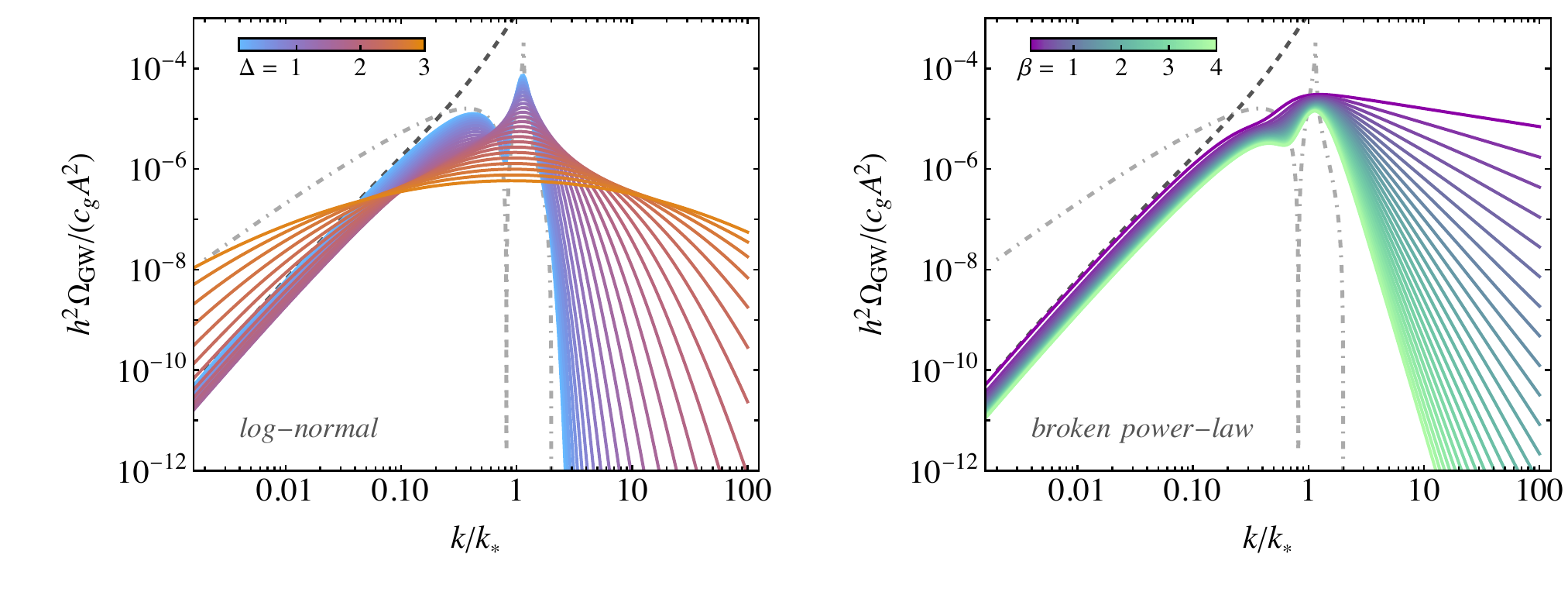}
  \caption{Examples of SIGW spectra induced by LN \emph{(left panel)} and BPL \emph{(right panel)} scalar curvature power spectra with $\alpha = 4$ and $\gamma = 1$ in a range of parameters. The dashed line shows the asymptotic tail~\eqref{eq:GW_asympt} for $\beta = 0.1$ and $\Delta = 0.1$. The dot-dashed line shows the SIGW spectrum for a monochromatic curvature power spectrum.
  }
  \label{fig:SIGW_examples}
\end{figure}

We report here some analytic expressions for the low $k$ asymptotics of GW spectra generated by an enhanced feature in the curvature fluctuations with BPL and LN shapes, i.e. Eqs.~\eqref{eq:PPL} and \eqref{eq:PLN} respectively (see also Refs.~\cite{Pi:2020otn,Yuan:2019wwo}). 
The corresponding SIGW spectra are shown in Fig.~\ref{fig:SIGW_examples}. Sufficiently far away from the peak, we can drop the $s$ dependence inside $P_\zeta$ and expand the transfer functions in Eq.~\eqref{eq:P_h_ts}
\bea\label{eq:GW_appr}
    \Omega_{\rm GW} (k)
    \stackrel{k\ll k_*}{\sim}  \frac{4 c_g \Omega_r}{5} k^3 \int_0^{\infty} \frac{\td q}{q^4} \,P(q)^2 \left[ \pi^2 + \ln^2\left( \frac{3 e^2}{4} \frac{k^2}{q^2}\right) \right]
    \propto k^3(1 + \tilde A \ln^2(k/\tilde k)), 
\eea
where $\tilde A$ and $\tilde k = \mathcal{O}(k_{*})$ are parameters that depend on the shape of the curvature power spectrum and $c_g \equiv g_*/g_*^0 \,\left(g_{*s}/g_{*s}^0\right)^{-4/3}$. For the power spectra \eqref{eq:PPL} and \eqref{eq:PLN}, the asymptotics can be obtained explicitly
\bea\label{eq:GW_asympt}
    \Omega^{\rm PL}_{\rm GW} (k)/k^3  
    &\sim \frac{4 c_g \Omega_r}{5 } 
    \frac{\gamma (\alpha/\beta + 1)^{2 \gamma}}{\alpha +\beta}
    \left(\frac{\beta }{\alpha }\right)^{-x_-} 
    \frac{\Gamma \left(x_+\right) \Gamma \left(x_-\right)}{\Gamma(x_- + x_+)}
    \bigg[\pi^2 
    + \frac{\gamma ^2}{(\alpha+\beta)^2}\left(\psi'\left(x_-\right)+\psi'\left(x_+\right)\right) \\
    & +  
    \left(\log \left(3k^2/4\right) + 2 + \frac{2\gamma}{\alpha +\beta } \left( \psi\left(x_+\right) - \psi\left(x_-\right) + \log \left(\alpha/\beta\right)\right) \right)^2\bigg] 
    \\
    \Omega^{\rm PL}_{\rm GW} (k)/k^3
    &\sim  \frac{2 c_g \Omega_r}{5\sqrt{\pi} \Delta}  e^{\frac{9 \Delta^2}{4}} 
    \left[\pi^2 + 2 \sigma^2 + \left(\log \left(3k^2/4\right)+2+3 \sigma ^2\right)^2\right]
\eea
where $x_- \equiv \gamma (2\alpha-3)/(\alpha+\beta)$, $x_+ \equiv \gamma (2 \beta+3)/(\alpha+\beta)$ and $\psi$ denotes the polygamma function. This tail fits the numerically evaluated SIGW tails well, as can be seen in Fig.~\ref{fig:SIGW_examples}.

At the $k \gg k_{*}$ tail, the SIGW spectrum tracks roughly $\mathcal{P}^2_{\zeta}(k)$ when it is sufficiently wide, that is, wider than the SIGW spectrum corresponding to the monochromatic curvature spectrum.

\begin{figure}[t]
  \centering
  \includegraphics[width=0.49\textwidth]{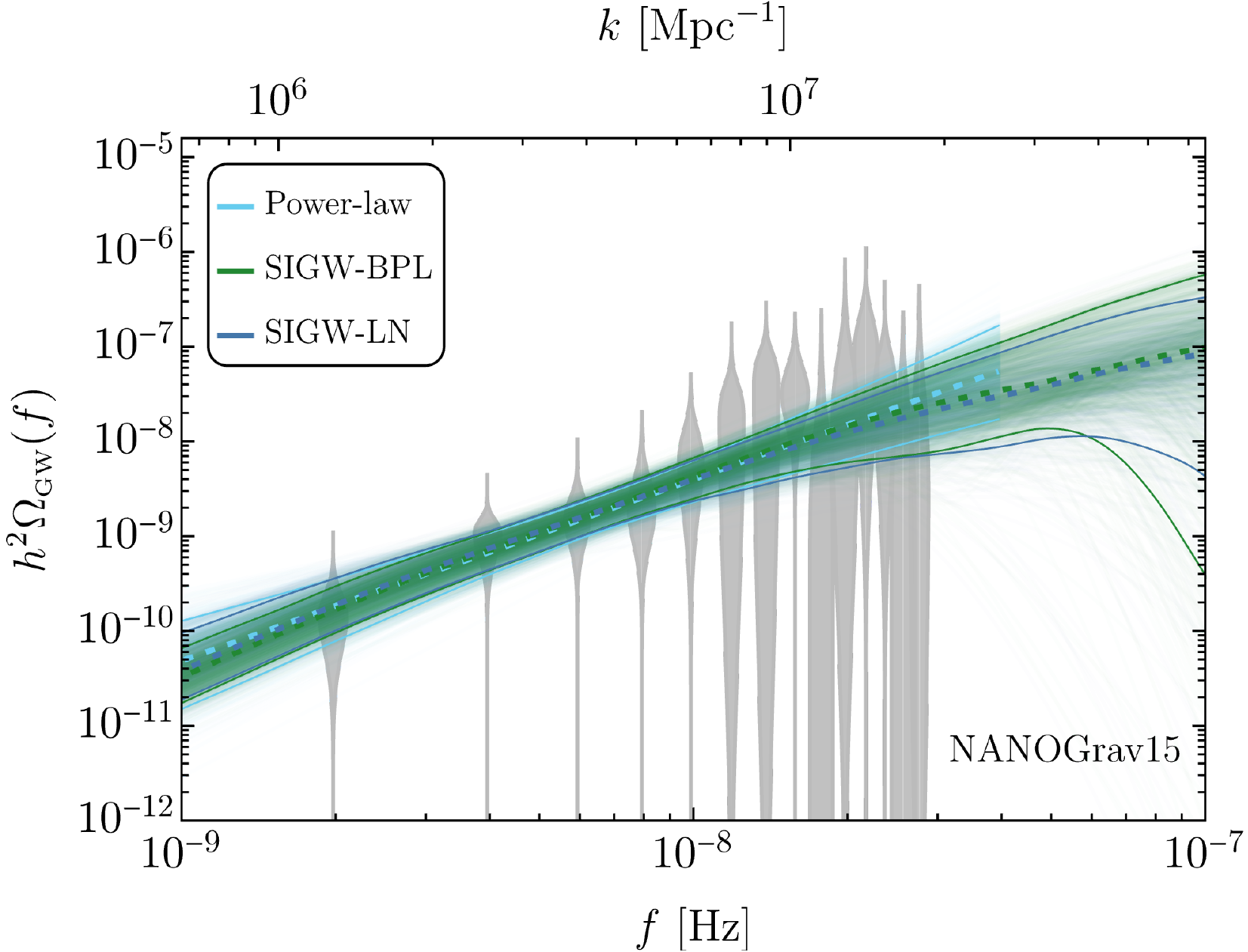}
  \includegraphics[width=0.49\textwidth]{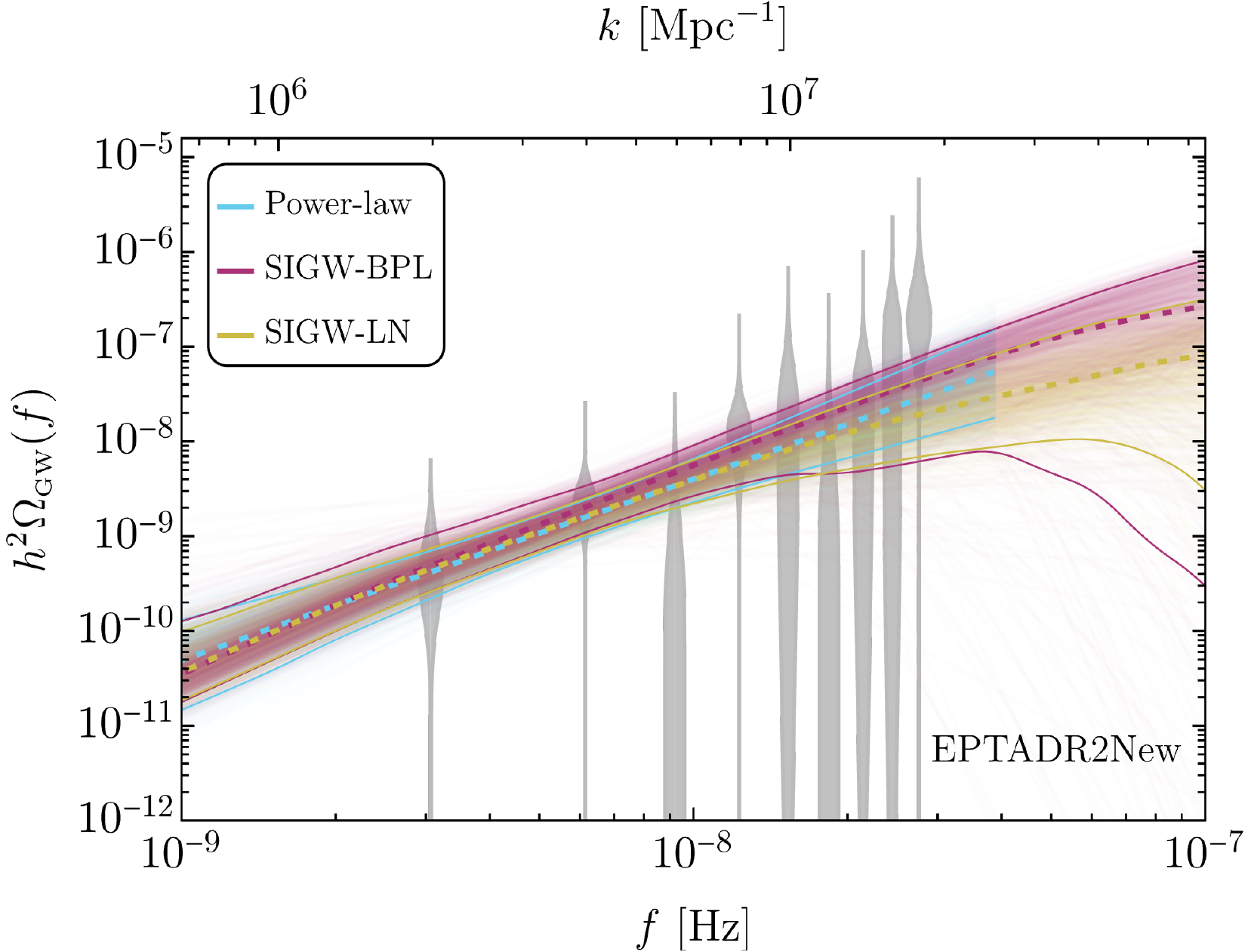}
  \caption{ 
SGWB spectra for the SIGW from BPL and LN models fitted to the NANOGrav15 data ({\it left panel}) and to the  EPTA data ({\it right panel}). 
For reference, we also show the SGWB power-law model used in~\cite{NANOGrav:2023gor, NANOGrav:2023hde}. 
The colored bands indicate the 90\%  C.I., while the gray posteriors show the NANOGrav15~\cite{NANOGrav:2023gor, NANOGrav:2023hde} and EPTA~\cite{EPTA:2023fyk, EPTA:2023sfo} data.}
  \label{fig:fits}
\end{figure}

In Fig.~\ref{fig:fits}, we show the best fit SGWB spectra for both the BPL and LN models compared to the NANOGrav15 (left panel) and EPTA (right panel) datasets.

\section{Dependency of the PBH formation parameters on the curvature power spectral shape}

The density contrast field $\delta$ averaged over a spherical region of areal radius $R_m\equiv a(t_H)r_me^{\zeta(r_m)}$,  and the compaction function $\mathcal{C}(r)$, defined as twice the local mass excess over the areal radius and evaluated at the scale $r_m$ at which it is maximized, are related by
\begin{align}\label{eq:Dicotomy}
\delta = \frac{1}{V_b(r_m,t_H)}
\int_{S^2_{R_m}}d\vec{x}\,
\delta\rho(\vec{x},t_H)
= \frac{\delta M(r_m,t_H)}{M_b(r_m,t_H)} 
=
\frac{2\left[M(r,t) - M_b(r,t)\right]}{R(r,t)} 
= \mathcal{C}\,.
\end{align}
As discussed in Ref.~\cite{Musco:2018rwt} and Refs.~therein, the gravitational collapse that triggers the formation of a PBH takes place
when the maximum of the compaction function $\mathcal{C}(r_m)$ is larger than a certain threshold value. Using Eq.\,(\ref{eq:Dicotomy}), we can relate threshold values in the compaction $\mathcal{C}_{\rm th}$ 
to the threshold for the density contrast $\delta_{\rm th}$.

In full generality, the threshold $\mathcal{C}_{\rm th}$ (or $\delta_{\rm th}$) depends on the shape of the collapsing overdensities, which is controlled by the curvature power spectrum~\cite{Germani:2018jgr, Musco:2018rwt, Musco:2020jjb}.
In this work, we follow Ref.~\cite{Musco:2020jjb}, where a prescription of how to compute the collapse parameter as a function of the curvature spectrum is derived. 
For completeness, we briefly report here the main steps to compute the threshold $\delta_{\rm th}$
and the shape parameter $\alpha_{\rm s}$. First of all, the maximum of the compaction function is located the the radius $r_m$, which can be found
solving numerically the integral equation
\begin{equation}
\int \frac{d k}{ k}\left[\left(k^2 {r}_m^2-1\right) 
\frac{\sin \left(k {r}_m\right)}{k {r}_m}+\cos \left(k {r}_m\right)\right]  P^{T}_\zeta(k) =0.
\end{equation}
Consequently, the shape parameter $\alpha_{\rm s}$ is obtained using 
\begin{equation}
F(\alpha_{\rm s})[1+F(\alpha_{\rm s})] \alpha_{\rm s}
=
-\frac{1}{2}\left[1+{r}_m \frac{\int d k k \cos \left(k {r}_m\right) P^{T}_\zeta(k)}{\int d k \sin \left(k {r}_m\right) P^{T}_\zeta(k)}\right],
\end{equation}
 where we introduced
 $  F(\alpha_{\rm s})
 =
 \left \{
 1-\frac{2}{5} e^{-1 / \alpha_{\rm s}} 
 \alpha_{\rm s}^{1-5 / 2 \alpha_{\rm s}}
 / 
 \left [
 \Gamma\left({5}/{2 \alpha_{\rm s}}\right)
 -\Gamma\left({5}/{2 \alpha_{\rm s}}, {1}/{\alpha_{\rm s}}
 \right)
 \right]
 \right \}^{1/2}
 $ to shorten the notation. 
Finally, once we determined shape parameter $\alpha_{\rm s}$, we can compute the threshold $\delta_{\rm th}$ using the relation \cite{Escriva:2019phb}
 \begin{equation}
\delta_{\rm th} \simeq \frac{4}{15} e^{-1 / \alpha_{\rm s}} \frac{\alpha_{\rm s}^{1-5 / 2 \alpha_{\rm s}}}{\Gamma\left(\frac{5}{2 \alpha_{\rm s}}\right)-\Gamma\left(\frac{5}{2 \alpha_{\rm s}}, \frac{1}{\alpha_{\rm s}}\right)}.
\end{equation}

\begin{figure}[h]
\begin{center}
$$\includegraphics[width=.48\textwidth]{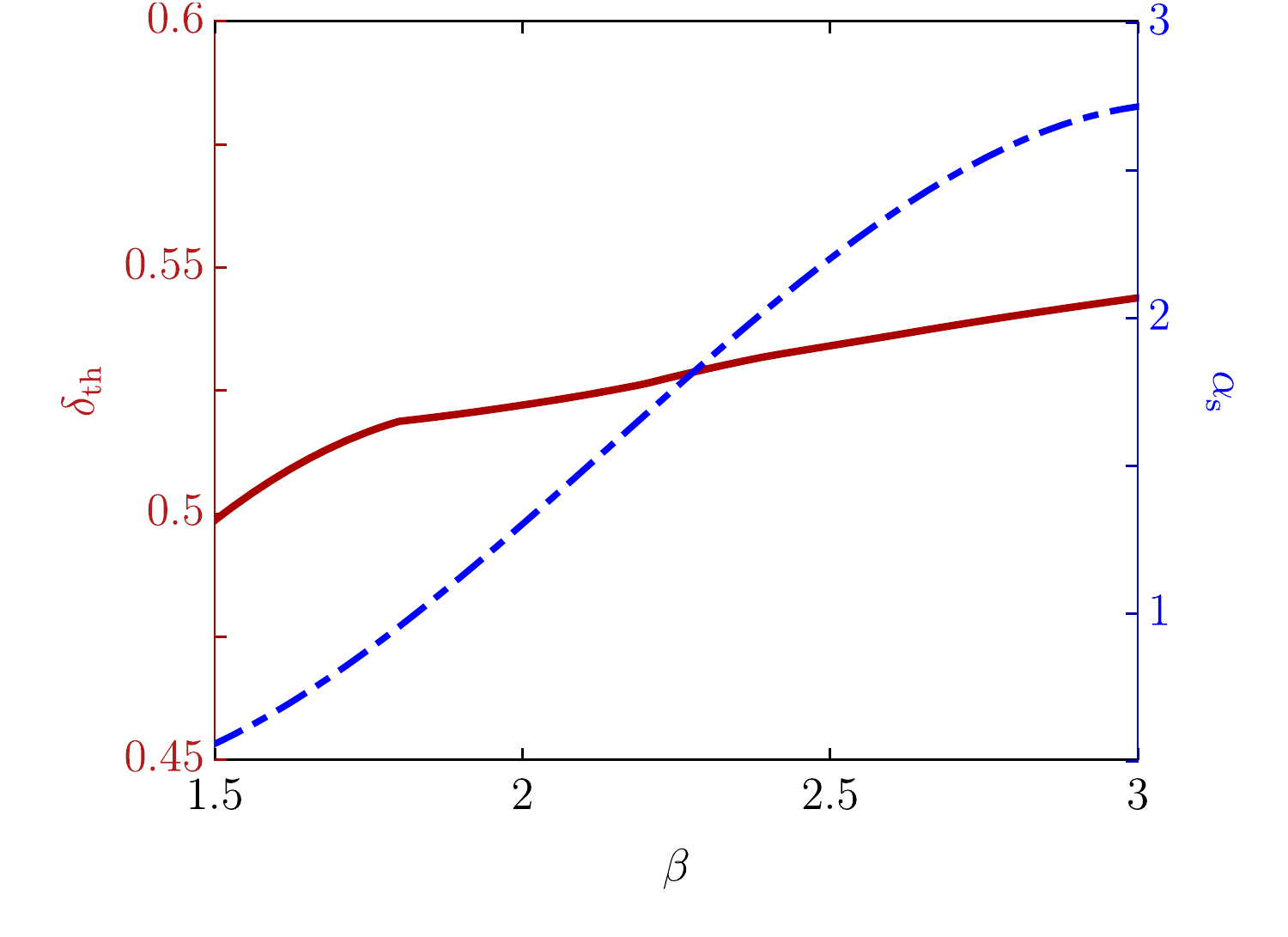}
\qquad\includegraphics[width=.48\textwidth]{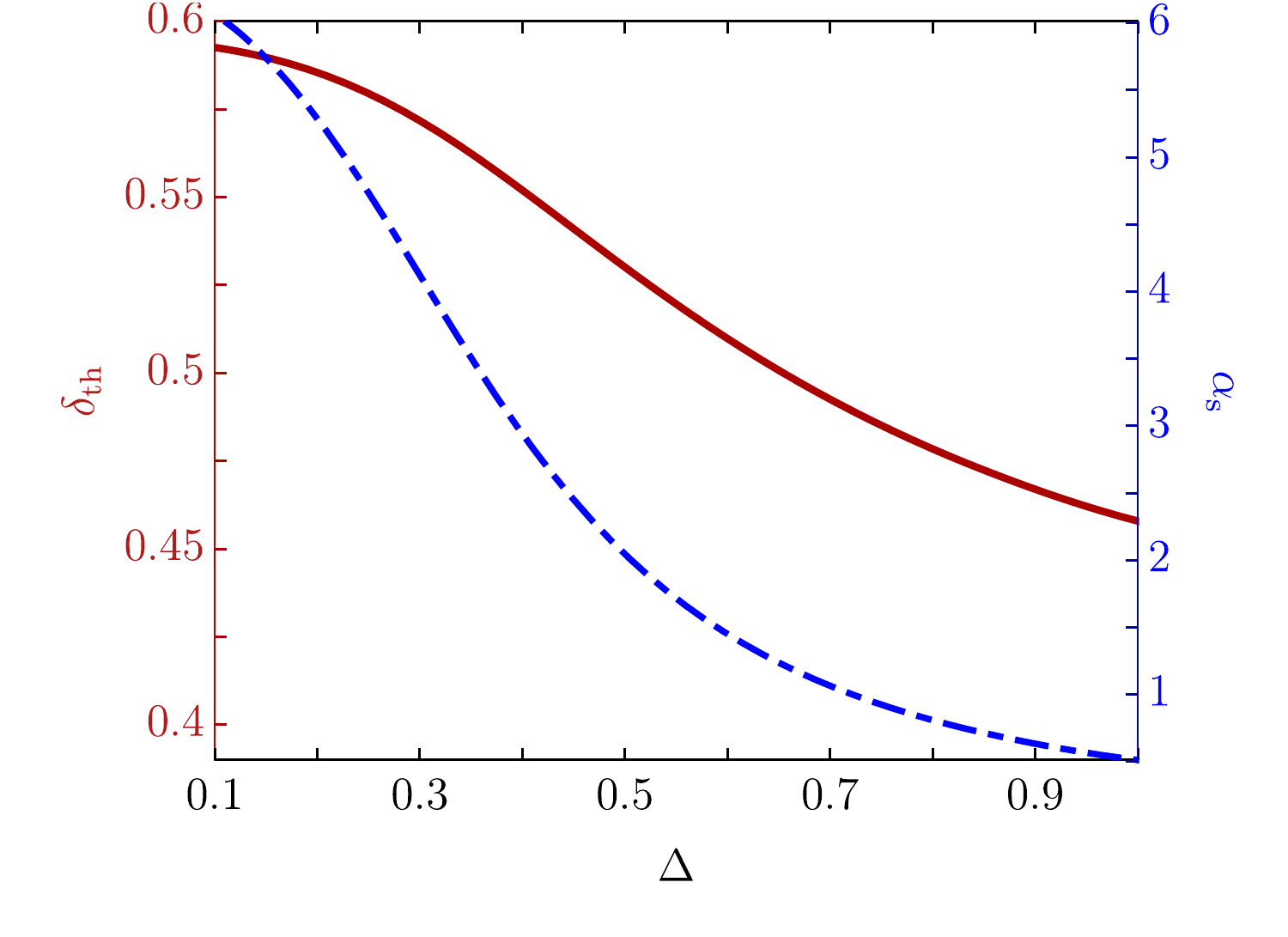}$$
\caption{ Threshold $\delta_{\rm th}$ and NG shape parameter $\alpha_{\rm s}$ for different shape of the power spectrum for a BPL \emph{(left panel)}, where we fix $\alpha=4$ and $\gamma=1$, and a log-normal \emph{(right panel)}.
 }\label{fig:ThShape}  
\end{center}
\end{figure}

In Fig.~\ref{fig:ThShape} we show how the threshold $\delta_{\rm th}$ and the shape parameter $\alpha_{\rm s}$ change respect the shape of the power spectrum for a BPL (right panel) and a LN (left panel).
It is worth emphasizing a general trend. As the spectrum becomes narrower, the threshold for collapse rises up to ${\cal O}({20})\%$ larger values. 
This is because, for broader spectra, more modes participate in the evolution of the collapsing overdensity, leading to a reduction of pressure gradients that facilitates the PBH formation, see Ref.~\cite{Musco:2020jjb} for more details.
It is important to note that the prescription outlined in Ref.~\cite{Musco:2020jjb} to compute the threshold for PBH collapse only accounts for NGs arising from the non-linear relation between the density contrast and the curvature perturbations. In principle, also primordial NGs beyond the quadratic approximation should be taken into account when computing the threshold value.
Following Refs.~\cite{Kehagias:2019eil,Escriva:2022pnz}, it appears that the effect on the threshold is small and at most of the order of a few percent. Interestingly, it follows the same impact NGs have on the statistics, i.e. negative (positive) NGs would tend to increase (decrease) the required amplitude of spectra. Therefore, neglecting this effect we are conservative. We left the inclusion of this effect for future work.

Due to a softening of the equation of state, the formation of PBH becomes more efficient during the QCD transition \cite{Jedamzik:1998hc, Byrnes:2018clq, Franciolini:2022tfm, Escriva:2022bwe, Musco:2023dak}. 
Consequently, the formation parameters change due to the softening of the equation of state. 
In our analysis presented in the main text, we computed $\alpha_{\rm s}$ for each power spectrum,
and used the results obtained in Ref.\cite{Musco:2023dak}, where  the formation parameters $\gamma(M_H), \mathcal{K}(M_H),\Phi(M_H)$ and $\mathcal{\delta}_{\rm th}(M_H)$ (or equivalently $\mathcal{C}_{\rm th}(M_H)(M_H)$) are obtained a function of $\alpha_s$.

\section{Primordial non-Gaussianities}\label{App:Curvaton}

\begin{figure}[h]
  \centering
  \includegraphics[width=0.325\textwidth]{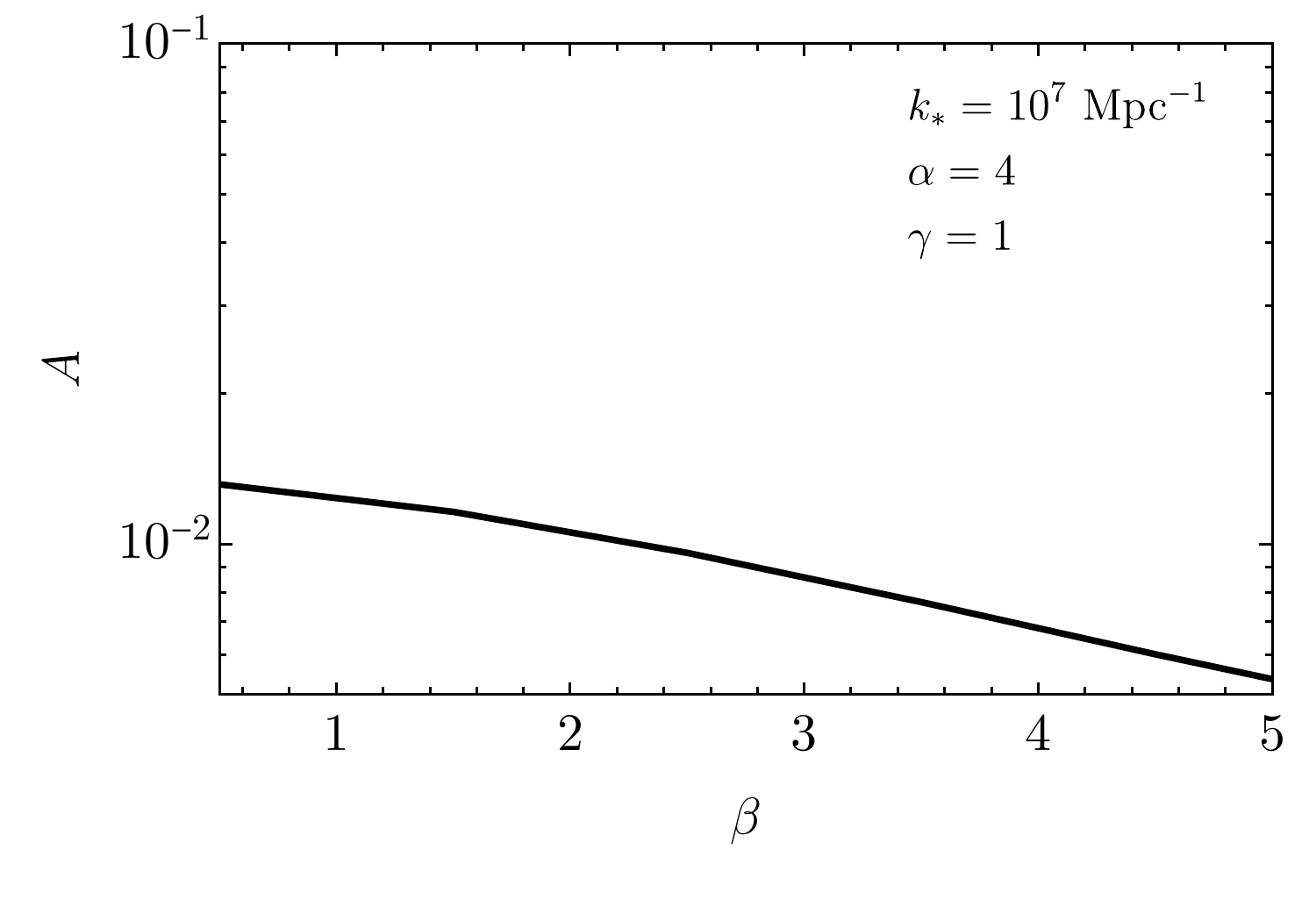}
  \includegraphics[width=0.325\textwidth]{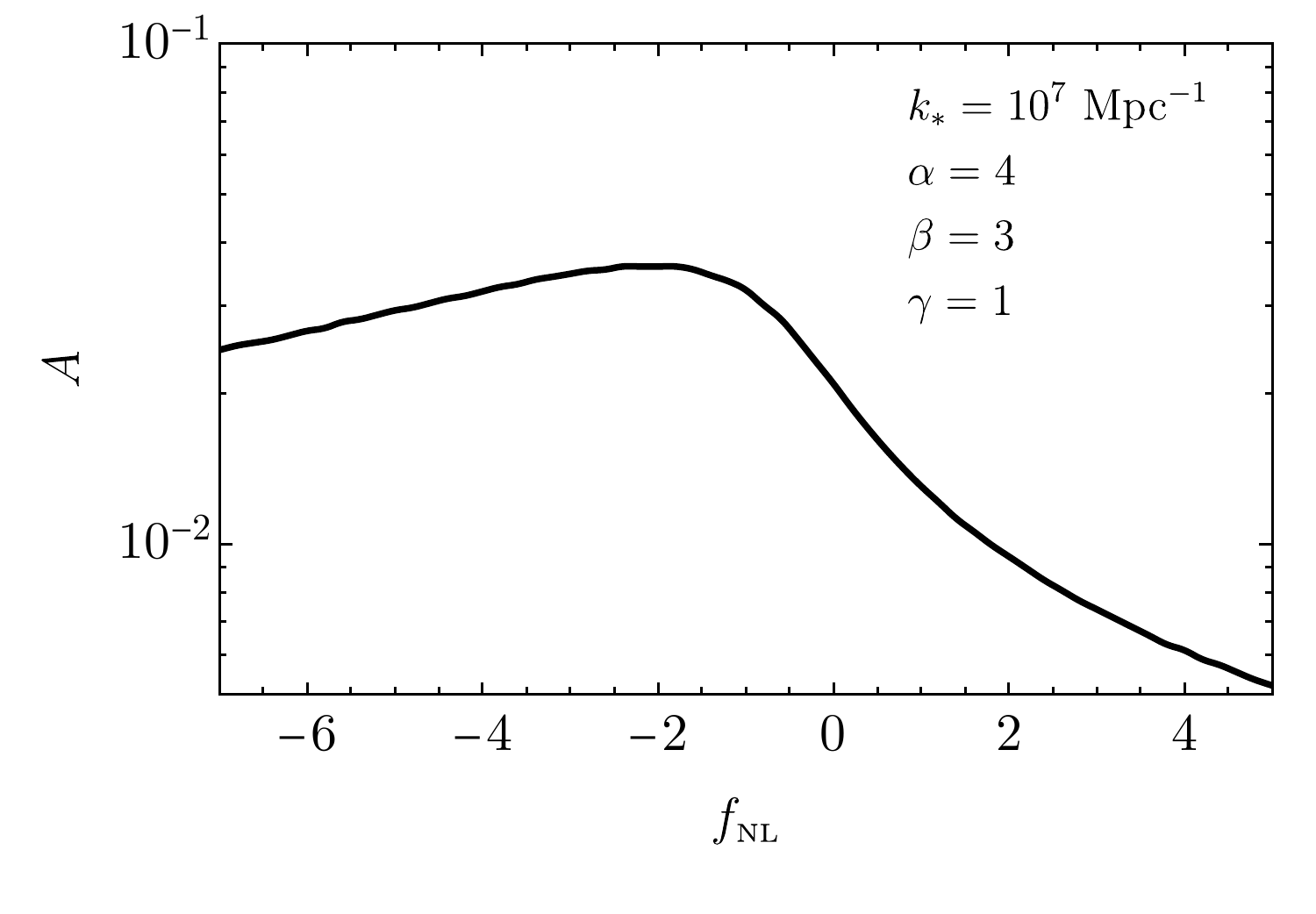}
  \includegraphics[width=0.325\textwidth]{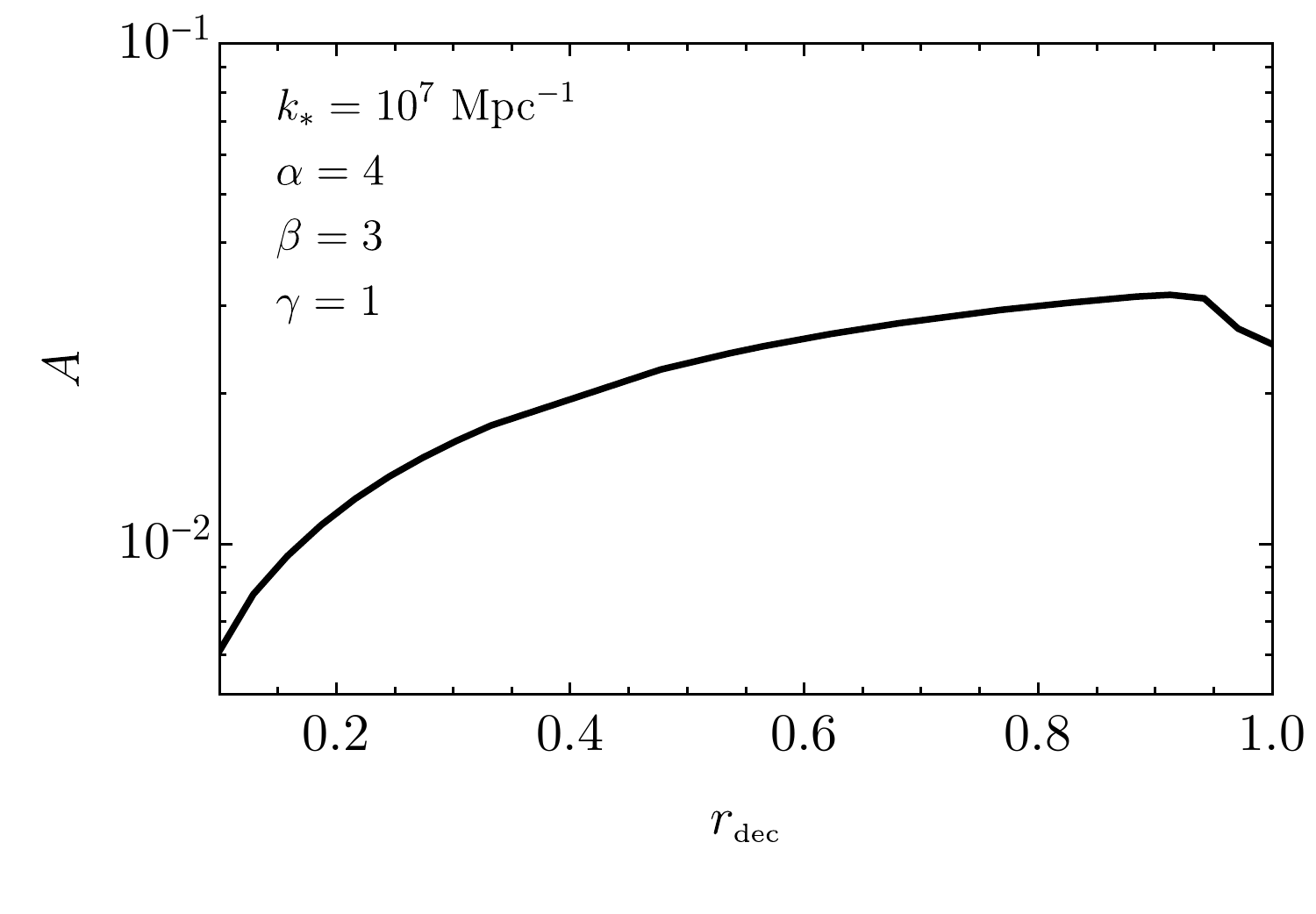}
  \caption{\emph{Left panel}: Amplitude values for a BPL curvature spectrum (Eq.~\eqref{eq:PPL}), fixing $\alpha=4$ and $\gamma=1$, in order to get $f_{\rm PBH}=1$ for quasi-inflection-point models with different $\beta$ values. 
  In order to isolate the effect of NGs, in this plot only we fix $\delta_{\rm th}=0.54$, as used in the other panels. 
  \emph{Middle panel}: Amplitudes required for $f_{\rm PBH}=1$ in the case of a BPL (Eq.~\eqref{eq:PPL}),  $\alpha=4$, $\beta=3$ and $\gamma=1$, in a model that only presents quadratic NGs (see  Eq.~\eqref{eq:FirstExpansion}) with different values of $f_{\rm NL}$. 
  \emph{Right panel}: The same as the middle panel but for the curvaton model (Eq.~\eqref{eq:MasterX}) as a function of $r_{\rm dec}$.}
  \label{fig:AppFNL}
\end{figure}

In this section, we discuss a few relevant details about the NG models considered in this work. 
We summarise the behaviour of the amplitude required to produce $f_\text{\tiny PBH }=1$ depending on the various types of NGs discussed in this paper in Fig.~\ref{fig:AppFNL}, assuming a reference BPL power spectrum.

\subsubsection{Quasi-inflection point} 
As described already in the main text, when presenting results inspired by the quasi-inflection point, the NG parameter $\beta$ is inherently determined by the UV slope of the power spectrum. As we can see from the left panel in Fig.~\ref{fig:AppFNL}, when one shrinks the shape of the power spectrum and keeps the threshold for collapse constant, we find that the amplitude should decrease to obtain $f_\text{\tiny PBH} = 1$. 
This is because NGs become larger with increasing $\beta$ in such models. 

One can also easily see that, by expanding Eq.~\eqref{eq:zeta_IP} at second order, 
$f_{\rm NL} = 5 \beta/12 $. 
Using the reference value $\beta = 3$ shown in Fig.~\ref{fig:abundance}, one finds $f_\text{NL} = 5/4$. As a consequence, we expect the NG correction to the SGWB to be below ${\cal O}(10^{-2})$, and no relevant correction from higher order terms in Eq.~\eqref{eq:P_h_ts} is expected.

\subsubsection{Quadratic non-Gaussianities} 

\begin{figure}[h]
\begin{center}
\includegraphics[width=.325\textwidth]{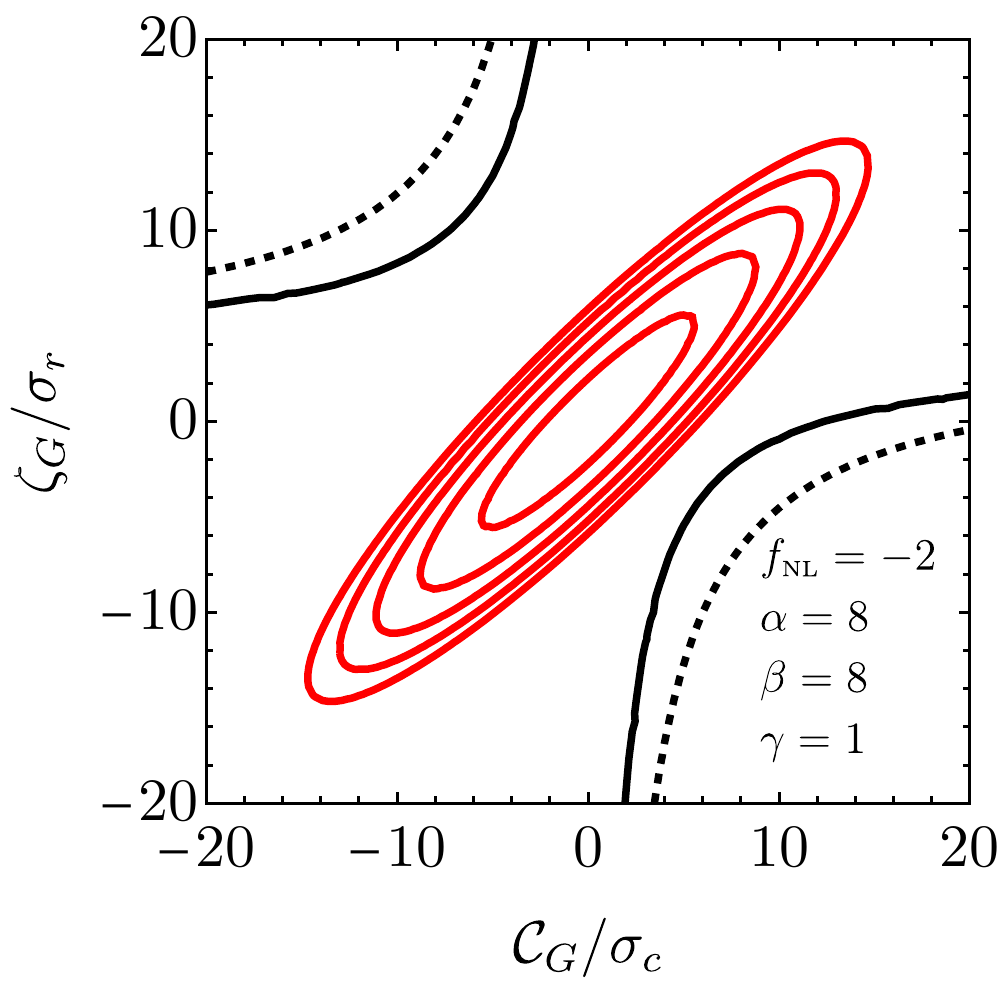}
\includegraphics[width=.325\textwidth]{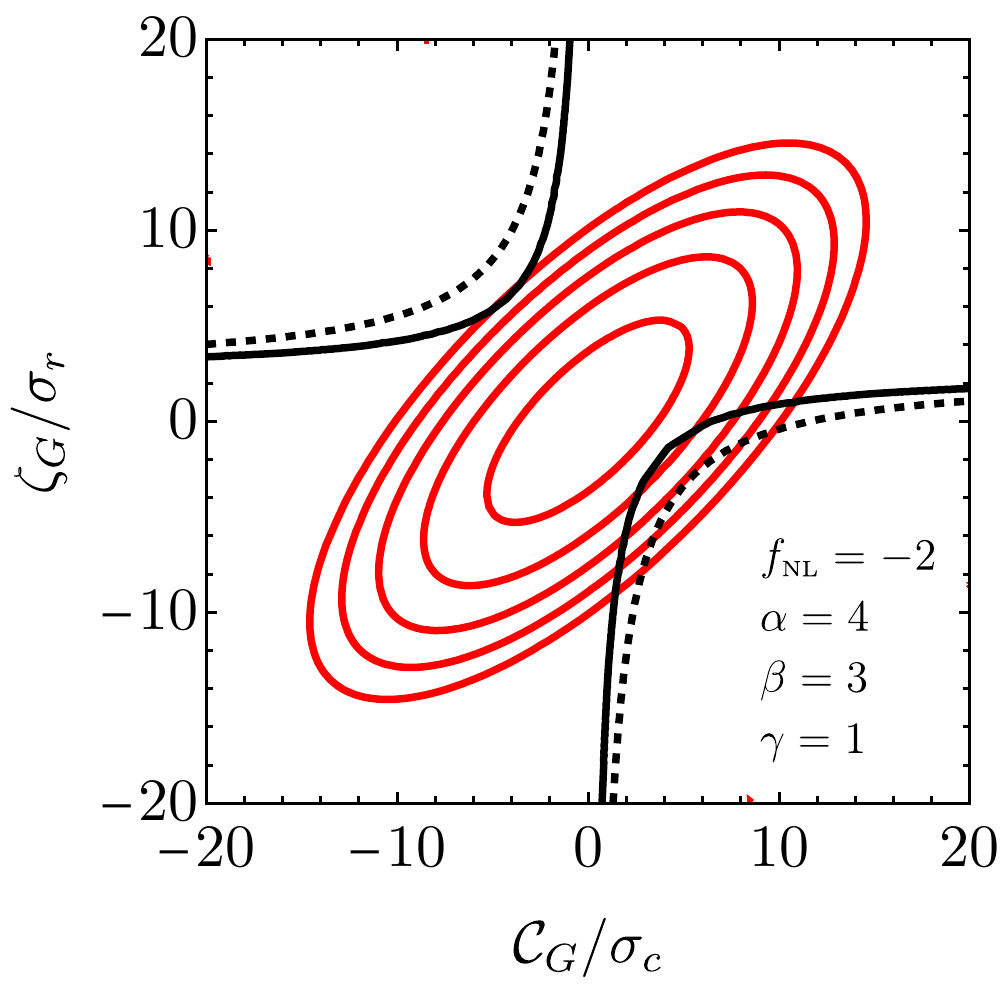}
\includegraphics[width=.325\textwidth]{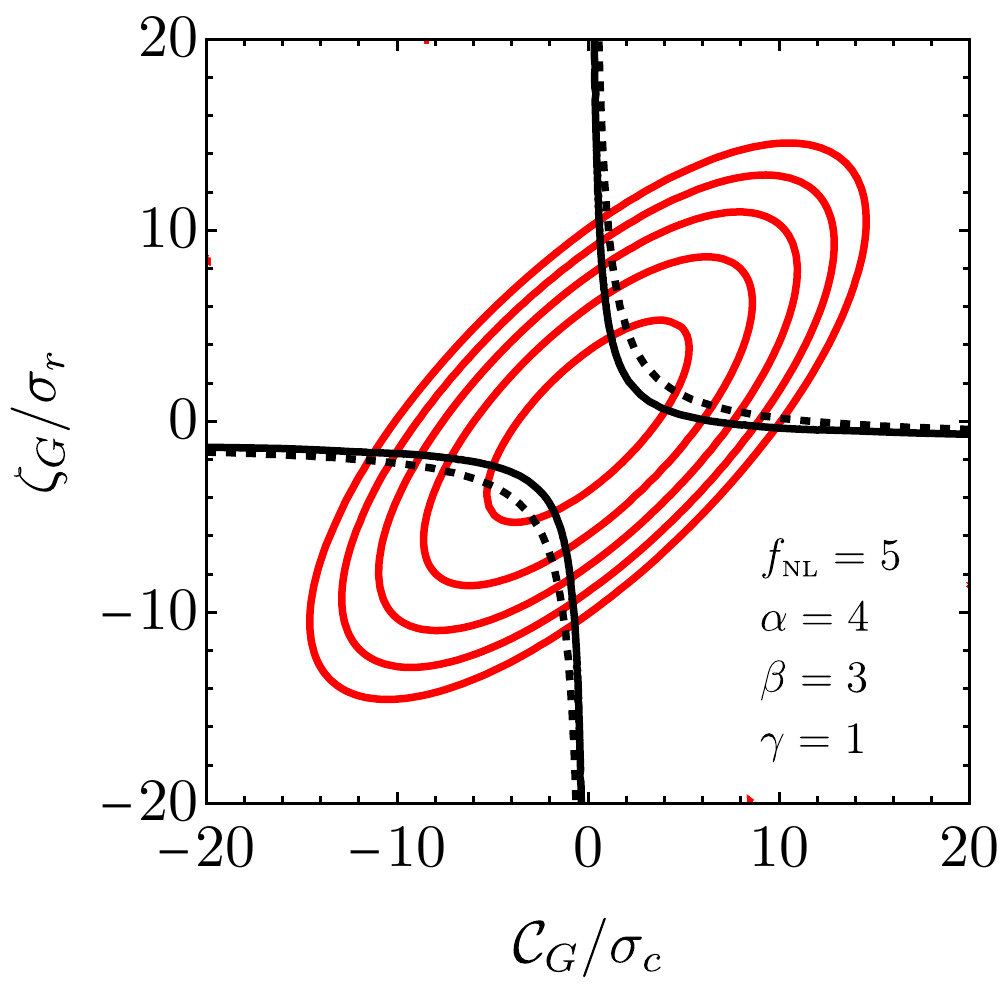}
\caption{ Two dimensional PDF as a function of $({\cal C}_\text{\tiny G}, \zeta_\text{\tiny G})$ compared to the over-threshold condition ${\cal C}>{\cal C}_\text{\tiny th}$.
In all panels, we considered the BPL power spectrum with an amplitude $A = 0.05$. The red lines indicates the contour lines corresponding to $\log_{10}(P_\text{\tiny G}) = {-45,-35,-25,-15,-5}$. The collapse of type-I PBHs takes place between the black solid and dashed lines (see more details in Ref.~\cite{Ferrante:2022mui}). 
\textit{Left panel:} Example of a very narrow power spectrum with $\alpha = \beta = 8$. The abundance is suppressed in the presence of negative $f_\text{NL} $ by the strong correlation between ${\cal C}_\text{G}$ and $\zeta_\text{G}$ obtained for narrow spectra.
\textit{Center panel:}
Example of negative non-Gaussianity and representative BPL spectrum. The PBH formation is sourced by regions of small $\zeta_\text{G}$ and positive ${\cal C}_\text{ G}$ or both negative ${\cal C}_\text{G}$ and $\zeta_\text{G}$.
\textit{Right panel:}
Example with positive $f_\text{NL}$, showing the region producing PBHs populates the correlated quadrants of the plot, at odds with that is found in the other panels. 
 }\label{fig:quadfnl}  
\end{center}
\end{figure}

Generically, one might mistakenly assume that by pushing the coefficient $f_{\rm NL}$ towards larger negative values, the required value of the power spectral amplitude associated with $f_\text{PBH} \simeq 1$ would subsequently rise. 
As we show in the middle plot of Fig.~\ref{fig:AppFNL}, however, a maximum amplitude is reached for $f_{\rm NL} \simeq -2$.
The reason for the appearance of such a peak and the subsequent decrease of A for $f_\text{NL}<-2$ can be understood as follows. 
When $f_\text{NL}$ is negative, one can still push the compaction function beyond the threshold, provided  $\zeta_\text{G}$ and ${\cal C}_\text{G}$ had opposite signs or one has small $\zeta_\text{G}$ and positive ${\cal C}_\text{G}$. 
In Fig.~\ref{fig:quadfnl}, we show the probability distribution for both Gaussian parameters $\zeta_\text{G}$ and ${\cal C}_\text{G}$ (red contours), compared to the overthreshold condition (between the black lines).
For realistic spectra, one always finds sufficient support of the PDF in the anti-correlated direction, and thus obtain a sizeable PBH abundance. Only in the limit of a very narrow power spectrum, one finds $\gamma_{cr}$ converging towards unity, which means a perfect correlation between $\zeta_\text{G}$ and ${\cal C}_\text{G}$, that leads to very small overlap with the parameter space producing overthreshold perturbation of the compaction function. 
This explains the appearance of a sharp rise of $A$ in the results Ref.~\cite{Young:2022phe} (see their Fig. 2), which, is however, expected only for extremely narrow spectra.

For this ansatz and with $f_{\rm NL}=-2$, we find that the amplitude of the power spectrum saturating the abundance of PBH is around $A \simeq 10^{-1.4}$, so we should expect a correction to the SIGW spectrum of the order $A(3/5f_{\rm NL})^2\simeq{\cal O}(0.05)$.
Still, this is subdominant compared to the leading order term used in this paper.

\subsubsection{Curvaton models} 
When presenting results inspired by the curvaton model, we will focus on primordial NG (derived analytically within the sudden-decay approximation~\cite{Sasaki:2006kq})
\be\label{eq:MasterX}
    \zeta = \log\big[X(r_{\rm dec},\zeta_{\rm G})\big]\,,
\ee
with
\begin{subequations}
\begin{align}\label{eq:XFunction}
    X &\equiv \frac{\sqrt{K}\left(1 + \sqrt{A K^{-\frac32}-1}\right)}{(3+r_{\rm dec})^{\frac13}}, 
    \\
    K & \equiv \frac{1}{2}\left((3+r_{\rm dec})^{\frac13}(r_{\rm dec}-1)P^{-\frac13} + P^{\frac13}\right), 
    \\
    P &\equiv A^2 + \sqrt{A^4 + (3+r_{\rm dec})(1-r_{\rm dec})^3}\,,
    \\
    A &\equiv \left(1 + \frac{3\zeta_{\rm G}}{2r_{\rm dec}} \right)^{2}r_{\rm dec}\,.
\end{align}
\end{subequations}
The parameter $r_{\rm dec}$ is the weighted fraction of the curvaton energy density  $\rho_{\phi}$ to the total energy density at the time of curvaton decay, defined by
\be
    r_{\rm dec} \equiv 
    \left.\frac{3 \rho_{\phi}}{3 \rho_{\phi} + 4 \rho_{\gamma}}\right|_{\rm curvaton\,\,decay}\,,
\ee
where $\rho_{\gamma}$ is the energy density stored in radiation after reheating. 
Thus, $r_{\rm dec}$ depends on the physical assumptions about the physics of the curvaton within a given model.

For comparison, the coefficients in the series expansion
$
    \zeta = \zeta_{\rm G} + (3/5)f_{\rm NL} \zeta_{\rm G}^2 + (3/5)^2g_{\rm NL} \zeta_{\rm G}^3 + \ldots
$
are given by
\be
    f_{\rm NL} = \frac{5}{3}\left(\frac{3}{4 r_{\rm dec}} - 1 - \frac{r_{\rm dec}}{2}\right), \qquad
    g_{\rm NL} = \frac{25}{54}\left(-\frac{9}{r_{\rm dec}} + \frac{1}{2} + 10r_{\rm dec} + 3r_{\rm dec}^2 \right)\,.
\ee
At $r_{\rm dec} \to 1$, the fluctuations arise from the curvaton field only, as it completely dominates the energy density budget at the time of decay. This gives
$
    \zeta = (2/3)\ln\left[1 + (3/2) \zeta_{\rm G}\right],
$
and $f_{\rm NL} = -5/4$, $g_{\rm NL} = 25/12$. Note that this mimics NG in inflection point models~\eqref{eq:zeta_IP} with an unphysical $\beta = -3$.
For the benchmark case we used in the main text, $r_{\rm dec}=0.9$, one finds $f_{\rm NL} = -1$, $g_{\rm NL} = 0.9$.
Also, we determine that the order of magnitude of the NG correction to the SIGW should be of order $A(3/5f_{\rm NL})^2\simeq {\cal O}(0.01)$, and we do not expect any relevant correction from higher order terms in Eq.~\eqref{eq:P_h_ts}.

\end{document}